\documentclass[a4paper,12pt]{article}
\usepackage{amsfonts,amsthm,amsmath,amssymb,graphicx,hyperref,youngtab,color,tcolorbox}
\numberwithin{equation}{section}
\usepackage{cite}
\usepackage{bbold}
\usepackage{amsmath}
\usepackage{bm}
\usepackage{young}

\begin{document}
\newtheorem{definition}{Definition}[section]
\newcommand{\be}{\begin{equation}}
\newcommand{\ee}{\end{equation}}
\newcommand{\bea}{\begin{eqnarray}}
\newcommand{\eea}{\end{eqnarray}}
\newcommand{\LE}{\left[}
\newcommand{\me}{\mathrm{e}}
\newcommand{\R}{\right]}
\newcommand{\nn}{\nonumber}
\newcommand{\Tr}{\text{Tr}}
\newcommand{\N}{\mathcal{N}}
\newcommand{\G}{\Gamma}
\newcommand{\vf}{\varphi}
\newcommand{\LL}{\mathcal{L}}
\newcommand{\Op}{\mathcal{O}}
\newcommand{\HH}{\mathcal{H}}
\newcommand{\arctanh}{\text{arctanh}}
\newcommand{\up}{\uparrow}
\newcommand{\down}{\downarrow}
\newcommand{\ket}[1]{\left| #1 \right>}
\newcommand{\bra}[1]{\left< #1 \right|}
\newcommand{\ketbra}[1]{\left|#1\right>\left<#1\right|}
\newcommand{\rd}{\partial}
\newcommand{\de}{\partial}
\newcommand{\ba}{\begin{eqnarray}}
\newcommand{\ea}{\end{eqnarray}}
\newcommand{\db}{\bar{\partial}}
\newcommand{\we}{\wedge}
\newcommand{\ca}{\mathcal}
\newcommand{\lr}{\leftrightarrow}
\newcommand{\f}{\frac}
\newcommand{\s}{\sqrt}
\newcommand{\vp}{\varphi}
\newcommand{\hvp}{\hat{\varphi}}
\newcommand{\tvp}{\tilde{\varphi}}
\newcommand{\tp}{\tilde{\phi}}
\newcommand{\ti}{\tilde}
\newcommand{\ap}{\alpha}
\newcommand{\pr}{\propto}
\newcommand{\mb}{\mathbf}
\newcommand{\ddd}{\cdot\cdot\cdot}
\newcommand{\no}{\nonumber \\}
\newcommand{\la}{\langle}
\newcommand{\lb}{\rangle}
\newcommand{\ep}{\epsilon}
 \def\we{\wedge}
 \def\lr{\leftrightarrow}
 \def\f {\frac}
 \def\ti{\tilde}
 \def\ap{\alpha}
 \def\pr{\propto}
 \def\mb{\mathbf}
 \def\ddd{\cdot\cdot\cdot}
 \def\no{\nonumber \\}
 \def\la{\langle}
 \def\lb{\rangle}
 \def\ep{\epsilon}

\title{\bf Backgrounds from Tensor Models: \\
A Proposal}
\author{Pablo Diaz\thanks{pablodiazbe@gmail.com}, \\
{\small \emph{Depatamento de Matem\'atica Aplicada, Universidad de Zaragoza.}}\\
}

\maketitle
\begin{abstract}
Although tensor models are serious candidates for a theory of quantum gravity a connection with classical spacetimes have been elusive so far. This paper aims to fill this gap by proposing a neat connection between tensor theory and Euclidean gravity at the classical level. The main departure from the usual approach is the use of Schur invariants (instead of monomial invariants) as manifold partners. Classical spacetime features  can be identified naturally on the tensor side in this new setup. A notion of locality is shown to emerge through Ward identities, where proximity between spacetime points translates into vicinity between Young diagram corners.  

\textbf{Keywords}: Tensor Models, Schur Operators, Backgrounds, Locality, Emergent Spacetime.
\end{abstract}
\newpage
\tableofcontents
\newpage
\section{Introduction}

One of the most fascinating challenges of physics nowadays is to understand the quantum nature of spacetime. In contrast to other theories which were developed in parallel with experiments, like quantum mechanics, the development of quantum gravity has to rely only on mathematical consistency due to the high energy experiments it would require to perform in order to test it in the lab. A necessary check for any quantum gravity theory is the recovery of Einstein gravity at the classical limit.

We still lack a unified framework for quantum gravity. However, several approaches have brought important insights from different perspectives. Known approaches to quantum gravity are string theory, non-commutative geometry, holography, spacetime triangulations, canonical quantum gravity, tensor theories... Probably one of them, if not a new one, will prevail in the future. It will probably be the one that offers a friendlier picture or permits more accurate calculations. As a comment, I must say that I do not find any conflict in the diversity of the current  spectrum of (sometimes overlapping) theories. At the end of the day, the success of a physical model to make predictions relies mostly on the consideration of the relevant degrees of freedom for the given phenomenon, and mathematics is rich enough to allocate those into separate frameworks.

An exciting feature of tensor theories is the idea of ``spacetime emergence''. Spacetime is not assumed {\it a priori}, but it is expected to appear combinatorially. Tensor models are expected to give a discretized (combinatorial) description of the Euclidean quantum gravity partition function
\begin{equation}\label{EQGPF}
    Z=\int dg \,e^{-I_E[g]}.
\end{equation}
The usual approach to quantum gravity from tensor models, and the line I will follow in this paper, is precisely the discretization of  \eqref{EQGPF}. However, there are recent developments, with interesting results, where by means of the connection between tensor and matrix theories, they apply holographic results and techniques to make contact with gravity, see for instance \cite{R4}. Maybe, these two
strategies, tensors as tools for discretization and holographic tensors, are not so different at the end of the day.  As an example, the $c=1$ string was motivated by summing over surfaces, but in the end there is an AdS/CFT like duality between matrix quantum mechanics and non-critical strings.

The idea of describing spacetime by means of tensor models comes historically from the remarkable success of matrix theories in describing 2-dimensional gravity \cite{earlierTM} at the sector where the matrix size, $N$, is large. However, the first tensor models that were proposed \cite{random1,random2,random3} were pathological at large $N$, and so the subject faded. In 2009, with the arrival of color tensor models \cite{color1,color2}, the situation changed. A well-defined $1/N$ expansion was found \cite{1N1,1N2,1N3}, the subject got revitalized \cite{GR,Diaz:2017kub,Diaz:2018xzt,Diaz:2018zbg,Diaz:2018eik,Itoyama:2019sos,Kemp:2019log,BR1,deMelloKoch:2017bvv,ito2,Itoyama:2018but,Delporte:2018iyf} and since then tensor models have become firm  candidates for a theory of quantum gravity. The interested reader can find a more comprehensive bibliographic information in \cite{Klebanov:2018fzb}, and the references therein.

  A precise connection between tensor invariants and piece-wise linear (PL) manifolds has been established. Invariants in tensor models are linear combinations of monomials  made of $n$ copies of a tensor $T$ which are contracted with $n$ copies of its complex conjugate $\overline{T}$ in a certain way.
  There is a natural map between these monomials and PL manifolds:  each pattern of contraction is interpreted as dictating how to glue simplices along their faces to build the manifold.   
   Monomial invariants are, so to say, the skeletons of PL manifolds. Remarkably enough, providing suitable identifications, it has been proven that the dynamics of tensor models reproduces the dynamics of triangulations driven by the Regge calculus \cite{Bonzom:2011zz}. Specifically, the amplitude of each tensor invariant appearing as a Wick contraction in the computation of an expectation value is associated with the amplitude of the corresponding triangulation related to the Regge action. This result together with the known fact that Regge discretizations lead to Einstein gravity at the continuum limit establishes a solid connection between tensor models and gravity. Let us remember that this connection involves some gauge fixing on the gravity side, since the triangulations must be equilateral.  
   
   Despite the success of the combinatorial description of PL manifolds, the classical limit and the description of backgrounds by tensor models has been elusive so far. The purpose of this paper is to fill this gap. The main point of departure of this paper from the usual setup is the consideration of a different set of invariants as partners of classical backgrounds. 
   Using representation theory arguments, a basis of invariants for any values of $n$ and $N$, the restricted Schur basis, has been found  \cite{Diaz:2017kub,Diaz:2018xzt,deMelloKoch:2017bvv,BR1}. There is a prominent set of invariants, the {\it Schur invariants}, which are easily constructed from characters of the symmetric group, and related to the restricted Schur basis by simple linear combinations.   Schur invariants are the candidates I propose for background partners. I will justify this choice in section \ref{SIAB} by showing how in the presence of a large Schur invariant the three-point function factorizes, meaning that any density matrix describing multi-particle states turns diagonal. 
   
   Schur invariants are linear combinations of monomials weighed by characters, they are labeled by $d$ Young diagrams with $n$ boxes each and a maximum of $N$ rows. 
   So, from the standard point of view (that is, with the identification of monomials with PL-manifolds), classical backgrounds (Schur invariants) occur in our setup as a collective behaviour of quantum contributions (monomial invariants). 
   
The first problem we face with the new set of invariants is that we do not know at first how to relate them to manifolds. Remember that as we leave monomials we lose the prescription to identify patterns of contraction with simplicial tilings. This is a central question and marks the starting point of the paper. In order to establish a connection between Schur invariants and manifolds I equate the partition functions of Euclidean gravity and tensor theory at a saddle point. This equation must be thought of as an ansatz. Using generic actions for both theories and some mathematical treatment, I obtain \eqref{eqatthepoint}, which tells us that, at the classical level, the curvature of the (discrete) manifold at each point is given by the expectation value of a corresponding Schur invariant. Furthermore, for Einstein gravity, the expectation value must be computed with the free tensor action. Notice that the connection between tensor theory and gravity we propose does not involve any gauge fixing.

A notion of locality arises in our setup via the Ward identities. Ward identities in tensor models involve the action of two operators, {\it cut} and {\it join}, defined in \eqref{CUT} and \eqref{JOIN}, respectively. Locality is linked to the cut action, which hits on invariants as a derivative.  The fact that the action of the cut operator  over Schur invariants produces all Schur invariants coming from the original with one box deleted in each label,  see \eqref{finalresult}, is crucial for a notion of locality in tensor models. It enable us to map the corners of the $d$ Young diagrams of a given Schur invariant to a grid of ``physical'' discrete spacetime points in a way that nearby points in the grid are nearby corners in the Young diagrams. 

Another insight we have from the use of Schur invariants in relation with classical backgrounds is that the limits $n,N\to \infty$ are not independent. This happens because the asymptotic Young diagrams must be limit shapes\cite{VK}, otherwise Schur invariants do not have a well-defined asymptotic limit\footnote{This is related to inductive definition and the representation theory of $S_{\infty}$. See the classical results by Thoma in \cite{Th} or a didactic review by Okounkov \cite{Ok}.}. This fact together with restriction of the number of rows in each diagram to be at most $N$, tells us that $n$ must grow as $N^2$.
One could wonder about the role (if any) the melonic sector will have in this picture. The answer is that for large invariants of size $n\sim N^2$ the melonic contribution to any expectation value is negligible\footnote{Here it applies the same discussion as in \cite{nonmelons} for invariants with $n\sim N$ (parallel to the original arguments in tensor models in \cite{Berkooz}), where it was shown that for such large invariants the contribution of non-melonic diagrams, for being so numerous, overwhelms the whole sum.}. Thus melon invariants, although leading for short invariants, play no role in our proposal for classical backgrounds\footnote{On the same lines, see \cite{R4} where, using holographic arguments, it is claimed that the melonic sector does not lead to an emergent geometry.}.        
Physical quantities in General Relativity should have a tensorial counterpart in the appropriate limit. If we think of asymptotically flat spacetimes, for simplicity, the ADM mass is a charge associated with the whole geometry. But, what is the ADM mass in the tensor world? Using general arguments, I claim that the ADM mass must be proportional to $n/N^2$, which is a fixed quantity for a given Schur invariant, and remains finite at $n,N\to \infty$. 
 
The paper is organized as follows. Section \ref{TM} provides  a brief introduction of tensor models: Invariants, the restricted Schur basis, Schur invariants, the action, the partition function and Gaussian correlators. It also sets the notation I will be using later. In section \ref{CwG}, I start by verifying the correspondence between large Schur invariants and classical backgrounds. I show how, in the presence of a large Schur invariant, three-point functions factorizes. This fact is interpreted as the large Schur invariant behaving as a classical background, where multi-particle states are seeing as independent excitations. Afterwards, in subsection \ref{PIA}, I put in contact both theories, tensor and gravity, by equating the respective partition functions at the saddle point in \eqref{eqPF}. The main result of the section is \eqref{eqfo}, where the on-shell Hilbert-Einstein action is computed by the Gaussian correlator of the corresponding tensor invariant. Locality is tackled in section \ref{Wid}. In this section it is shown how the use of the Ward identities in tensor models permits us to write Gaussian correlators as correlators involving the cut operator as in \eqref{Wardgaussian}. This is crucial for the emergence of a notion of locality in tensor models: hitting as a derivative, the cut operator acts on Schur invariants deleting a corner box in each of the invariant's labels. Those distinguished corners can be mapped to the grid which, on the gravity side, is discretized space. The most important result of section \ref{Wid}, and perhaps of the paper, is the equation \eqref{eqatthepoint}. Finally, in section \ref{BH}, I propose a tensor quantity which seems reasonable to relate to the ADM mass of spacetimes.

\section{Tensor models}\label{TM}
In this section, I review known facts of tensor models and set the notation. I also define the invariants which will play a role in the partition function and will be relevant in this paper. More information can be found in the appendices and in the references provided. 

\subsection{Invariants}

The basic object of color tensor models is the tensor $T$ of order $d$  and size $N$. The tensor $T$ is a box of $N^d$ complex numbers whose components transform under the gauge group $U(N)^{\times d}$ as, 
\begin{equation}
 T_{j_1j_2\dots j_d}=\sum_{i_1,\dots, i_d}U_1(N)_{j_1}^{i_1} \cdots U_d(N)_{j_d}^{i_d}  T_{i_1i_2\dots i_d},
\end{equation}
where, with $U_i(N)$, I am emphasizing that each component transforms under a different copy of $U(N)$.
The complex conjugate is a contravariant tensor that transforms as
\begin{equation}
 \overline{T}^{j_1j_2\dots j_d}=\sum_{i_1,\dots, i_d}\overline{U}_1(N)^{j_1}_{i_1} \cdots \overline{U}_d(N)^{j_d}_{i_d}  \bar{T}^{i_1i_2\dots i_d}.
\end{equation}
Invariants under $U(N)^{\times d}$ are made of $n$ copies of $T$ and $n$ copies of $\overline{T}$ as we contract all the indices of the tensors by pairs $(T,\overline{T})$ respecting the index position. So, first indices only contract with first indices, and so on. Any possible invariant can obtain as a linear combination of the elements of the set
\begin{equation}\label{spanset}
\Big\{\mathcal{O}_{\alpha_1\dots\alpha_d}=\prod_{p=1}^nT_{i_1^pi_2^p\dots i_d^p}\overline{T}^{i_1^{\alpha_1(p)}i_2^{\alpha_2(p)}\dots i_d^{\alpha_d(p)}}|\,(\alpha_1\dots\alpha_d)\in S_n^d\Big\},
\end{equation}
where subscripts and superscripts have been assigned to indices in order  to specify the location of the component and the slot each tensor occupies in the string of $n$ copies. As the notation in tensor theory quickly proliferates, for practical reasons I will reduce it as much as possible by omitting indices whenever they are not strictly necessary. The set \eqref{spanset} has often been called {\it permutation basis} in the literature, see \cite{GR}.

\subsection{Notation}
Besides the usual notation for tensor models, I will adopt a vector notation for $d$-tuples
\begin{equation}
    \vec{\alpha}=(\alpha_1,\dots,\alpha_d),\quad \alpha_i\in S_n, \qquad \vec{\mu}=(\mu_1,\dots,\mu_d),\quad \mu_i\vdash n.
\end{equation}
The product of two elements of $S_n^d$, and  a diagonal product of and element of $S_n$ with an element of $S_n^d$ will be written as
\begin{equation}
   \vec{\alpha}\cdot \vec{\beta}=(\alpha_1\beta_1,\dots,  \alpha_d\beta_d)\quad \text{and}\quad \vec{\alpha}\cdot \sigma=(\alpha_1\sigma,\dots,  \alpha_d\sigma),
\end{equation}
respectively. This vector notation applies to every mathematical object found in this paper. Thus,
for a product of characters I will use the notation
\begin{equation}
    \chi_{\vec{\mu}}(\vec{\alpha})=\chi_{\mu_1}(\alpha_1)\cdots \chi_{\mu_d}(\alpha_d),
\end{equation}
and for dimensions of the symmetric group and the unitary group I will, respectively, write
\begin{equation}
    d_{\vec{\mu}}=d_{\mu_1}\cdots d_{\mu_1} \quad \text{and}\quad \text{Dim}_{\vec{\mu}}(N)=\text{Dim}_{\mu_1}(N)\cdots \text{Dim}_{\mu_d}(N).
\end{equation}
Quantities like the Kronecker coefficients will be expressed as
\begin{equation}
    g_{\vec{\mu}}=g_{\mu_1\dots\mu_d}.
\end{equation}
Trace and Schur invariants will be denoted as
\begin{equation}
    \mathcal{O}_{\vec{\alpha}}=\mathcal{O}_{\alpha_1\dots \alpha_d}, \qquad \mathcal{O}_{\vec{\mu};ij}= \mathcal{O}_{\mu_1\dots\mu_d;ij}.
\end{equation}

In this paper, I will generally use prime Greek letter to mean objects related to $n-1$ elements. I find it especially convenient when dealing with the ``cut'' operation. There, I will call $\alpha'$ the permutation of $n-1$ elements that results from $\alpha$ after deleting the letter ``n''. The same logic will be used for Young diagrams, where the notation $\mu'\nearrow \mu$ means that the diagram $\mu'\vdash n-1$ is obtained from $\mu$ after deleting one corner box. This notation will be extended vectorially, thus
\begin{equation}
    \vec{\mu'}\nearrow \vec{\mu},
\end{equation}
will refer to a $d$-tuple of diagrams $\vec{\mu'}$ that is obtained from the $d$-tuple $\vec{\mu}$ after deleting
one corner box in each $\mu_i$.

\subsection{Restricted Schur basis and Schur invariants}
Despite the name, the elements of the set \eqref{spanset}, although they span the space of invariants, do not form a basis. They overexpress the space of invariants. This is not merely because of the obvious equivalence 
\begin{equation}
 \mathcal{O}_{\tau\cdot\vec{\alpha}\cdot \sigma}=\mathcal{O}_{\vec{\alpha}},   
\end{equation}
a redundancy that may be removed by considering only double coset representatives, but also because the elements of \eqref{spanset} are not linearly independent for $n>N$. 
Using arguments of representation theory, the exact number of invariants for given $N$ and $n$ was found\footnote{See also \cite{Ben1,Ben2} for the counting of invariants in tensor theories with orthogonal gauge group.} \cite{Diaz:2017kub,deMelloKoch:2017bvv,BR1,Diaz:2018xzt} and the natural basis adapted to the counting, the {\it restricted Schur basis}, was constructed\footnote{Analogous bases of operators have been constructed in matrix models. Firstly, in \cite{CJR} for a single matrix model, and later \cite{R1} and \cite{R2} for multimatrix models. See also, \cite{Ram1,Ram2} for other multimatrix model bases.}, see appendix \ref{RSB} for details. It is the set
\begin{equation}
   \{\mathcal{O}_{\vec{\mu};ij}|\quad \mu_i\vdash n,\quad l(\mu_i)\leq N, \quad i,j=1,\dots, g_{\vec{\mu}}\}.   
\end{equation}
The condition $l(\mu_i)\leq N$ forces each Young diagram $\mu_i$ to have a maximum of $N$ rows. 
The elements of the basis can be written as
\begin{equation}
\mathcal{O}_{\vec{\mu};ij}=\sum_{\vec{\alpha}\in S_n^d}F_{\vec{\mu};ij}(\vec{\alpha})\mathcal{O}_{\vec{\alpha}},
\end{equation}
for suitable complex double coset invariant functions $F_{\vec{\mu};ij}(\vec{\alpha})$ which fulfill the
convolution algebra
\begin{equation}
  F_{\vec{\mu};ij}*F_{\vec{\nu};kl}(\vec{\alpha})=\delta_{\vec{\mu}\vec{\nu}}\delta_{jk}F_{\vec{\mu};il}(\vec{\alpha}).  
\end{equation}
The functions $F_{\vec{\mu};ij}(\vec{\alpha})$ are projectors in the labels $\vec{\mu}$ and intertwiners in the labels $ij$.  \\
For reasons that will become clear later, in this paper we will be interested in a subset of invariants which we will call {\it Schur invariants}\footnote{Also called permutation centralizer algebras in the literature, see \cite{MR}.}, and are defined as
\begin{equation}\label{defschur}
 \mathcal{O}_{\vec{\mu}}\equiv \sum_{i=1}^{g_{\vec{\mu}}} \mathcal{O}_{\vec{\mu};ij}.  
\end{equation}
Schur invariants are build out of projectors and can be constructed explicitly. They are
\begin{equation}\label{defschur1}
 \mathcal{O}_{\vec{\mu}}= \frac{1}{n!^{d-1}}\sum_{\vec{\alpha}\in S_n^d}d_{\vec{\mu}}\chi_{\vec{\mu}}(\vec{\alpha})\mathcal{O}_{\vec{\alpha}},  
\end{equation}
see appendix \ref{SO} for details.

\subsection{Action, partition function and correlators}
The partition function of the theory is
\begin{equation}\label{pftm}
    Z[\lambda]=\int dT\,d\overline{T}\exp\Big(-\frac{N^{d-1}}{2}S[T,\overline{T}]\Big).
\end{equation}
The letter $\lambda$ encodes the couplings of all the interacting terms. The factor $N^{d-1}$ in front of the action makes the model asymptotically free as $N\to\infty$ \cite{G1}.\\
The most general action, which includes all the invariant operators of the theory, is
\begin{equation}\label{TMA}
     S[T,\overline{T}]= T\cdot\overline{T}+\frac{1}{N^{d-1}}\sum_{\vec{\mu},ij}\lambda_{\vec{\mu},ij}\mathcal{O}_{\vec{\mu},ij}.
 \end{equation}
 In this paper, we will be considering the sector of Schur invariants, so the action we will use is
 \begin{equation}\label{SA}
     S[T,\overline{T}]= T\cdot\overline{T}+\frac{1}{N^{d-1}}\sum_{\vec{\mu}}\lambda_{\vec{\mu}}\mathcal{O}_{\vec{\mu}}.
 \end{equation}
With \eqref{pftm}, the two-point function for the free theory of single tensors reads 
\begin{equation}
    \langle T_{i_1\dots i_d} \, \overline{T}^{j_1 \dots j_d}\rangle_0=\frac{1}{N^{d-1}}\delta_{i_1}^{j_1}\cdots\delta_{i_d}^{j_d},
\end{equation}
where the subscript ``0''  indicates that is a Gaussian average, no subscript meaning that the average involves the full action \eqref{SA}.
The correlator of the trace operators made of $2n$ tensors are
\begin{equation}\label{corretr}
    \langle \mathcal{O}_{\vec{\alpha}}\rangle_0= \frac{1}{N^{n(d-1)}}\sum_{\sigma\in S_n}N^{C(\vec{\alpha}\cdot \sigma)}.
\end{equation}
The Gaussian average of Schur operators $\mathcal{O}_{\vec{\mu}}$ are  computed in appendix \ref{RSB}. They are
\begin{equation}
    \langle \mathcal{O}_{\vec{\mu}}\rangle_0 = \frac{n!^d}{N^{n(d-1)}} \frac{\text{Dim}_{\vec{\mu}}(N)}{d_{\vec{\mu}}}g_{\vec{\mu}}=\frac{1}{N^{n(d-1)}} f_{\vec{\mu}}(N)g_{\vec{\mu}}.
\end{equation}

\section{Connection with gravity at the saddle point} \label{CwG}
Via triangulations it has been established a close relation between tensor models and gravity. Every invariant built on $2n$ $(d+1)-$tensors can be associated with a triangulation with $2n$ $d$-simplicies where the pattern of contraction of the indices encode the details of the triangulation. The Wick contractions of observables built on $2n$ tensor of order $d$ can be interpreted as invariants of order $d+1$, which may be associated to triangulations made of $2n$ $d$-simplices. Remarkably enough, it has been proven \cite{Bonzom:2011zz} that, provided the appropriate identifications, the statistics of tensor models match those of triangulations driven by Regge calculus, what establishes a solid connection between tensor models and gravity at the level of partition functions.  

It is not easy to find the tensor sector that corresponds to the continuum limit for gravity, that is, classical gravity. It will happen at large $N$, and there is a common belief that it should be at a fixed point of a certain renormalization flow, see \cite{astrid} and the references therein. This way, classical gravity would be sitting at a universality class where any detail of a specific triangulation would be irrelevant, as it should be. Then, we recover the necessary symmetry under diffeomorphisms of classical gravity\footnote{Different triangulations reduce to different coordinate systems at the continuum limit of Regge calculus.}. The big question is if tensor invariants could also, in some large limit of $n$ and $N$, encode a background, a classical solution of gravity. The main goal of this paper is to propose a collections of invariants (Schur invariants) which can be associated to backgrounds, as well as to establish a precise relation between them.

\subsection{Schur invariants and backgrounds}\label{SIAB}

I claim that Schur invariants correspond to backgrounds. This claim is supported by the factorization of the three-point function at large $n,N$ in the presence of a Schur invariant. Let us define
\begin{equation}
    \langle\langle \mathcal{O}_{\vec{\alpha}} \rangle\rangle_{\vec{\mu}}\equiv \frac{ \langle \mathcal{O}_{\vec{\alpha}}\mathcal{O}_{\vec{\mu}}\rangle}{\langle \mathcal{O}_{\vec{\mu}}\rangle},
\end{equation}
where $\vec{\mu}$ labels a Schur invariant with $\mu_i\vdash n$, and $\mathcal{O}_{\vec{\alpha}}$ is an invariant made of a few (order 1) tensors. Thus, $\mathcal{O}_{\vec{\alpha}}$ will be interpreted as an excitation of the background $\mathcal{O}_{\vec{\mu}}$. 
For the classical behaviour of the background, we need to prove that for large $n,N$,
\begin{equation}\label{factor}
  \langle\langle \mathcal{O}_{\vec{\alpha}} \mathcal{O}_{\vec{\beta}} \rangle\rangle_{\vec{\mu}} \approx \langle\langle \mathcal{O}_{\vec{\alpha}}  \rangle\rangle_{\vec{\mu}}\, \langle\langle  \mathcal{O}_{\vec{\beta}} \rangle\rangle_{\vec{\mu}}.
\end{equation}
Condition \eqref{factor} assures the independence of the states $\mathcal{O}_{\vec{\alpha}}$ and $\mathcal{O}_{\vec{\beta}}$ when happening in the large ``environment'' $\mathcal{O}_{\vec{\mu}}$. Consequently, the density matrix of any ``multiparticle'' state turns diagonal. For this reason, I will assume that if the condition \eqref{factor} holds for any $\mathcal{O}_{\vec{\alpha}}$ and $\mathcal{O}_{\vec{\beta}}$ then $\mathcal{O}_{\vec{\mu}}$ is a background, and $\mathcal{O}_{\vec{\alpha}}$ and $\mathcal{O}_{\vec{\beta}}$ should be thought of as excitations of $\mathcal{O}_{\vec{\mu}}$.\\    
For simplicity I am going to consider $\mathcal{O}_{\vec{\alpha}}$ and $\mathcal{O}_{\vec{\beta}}$ also Schur invariants with a  number of tensor copies $n_1$ and $n_2$, respectively, where $n_1,n_2\ll n$. So, one should think of $\mathcal{O}_{\vec{\alpha}}$ and $\mathcal{O}_{\vec{\beta}}$ as (Schur) excitations of $\mathcal{O}_{\vec{\mu}}$. \\
With the definition of Schur invariants \eqref{defschur1}, and applying \eqref{productofSchurs} for the product of two Schur invariants, we have
\begin{equation}\label{entangled}
    \langle \mathcal{O}_{\vec{\alpha}} \mathcal{O}_{\vec{\beta}} \mathcal{O}_{\vec{\mu}} \rangle=\frac{n!n_1!n_2!}{(n+n_1+n_2)!}\frac{1}{N^{(n+n_1+n_2)(d-1)}}\sum_{\lambda_i\vdash n+n_1+n_2}\frac{d_{\vec{\alpha}}d_{\vec{\beta}}d_{\vec{\mu}}}{d_{\vec{\lambda}}}f_{\vec{\lambda}}\,g_{\vec{\lambda}}\, C^{\vec{\lambda}}_{\vec{\mu}\vec{\alpha}\vec{\beta}},
\end{equation}
and 
\begin{equation}\label{factorized}
 \small{\frac{ \langle \mathcal{O}_{\vec{\alpha}}\mathcal{O}_{\vec{\mu}}\rangle \langle \mathcal{O}_{\vec{\beta}}\mathcal{O}_{\vec{\mu}}\rangle}{\langle \mathcal{O}_{\vec{\mu}}\rangle}=\frac{n!^2n_1!n_2!}{(n+n_1)!(n+n_2)!}\frac{1}{N^{(n+n_1+n_2)(d-1)}}\sum_{\substack{\lambda'_i\vdash n+n_1\\
 \lambda''_i\vdash n+n_2}} \frac{d_{\vec{\alpha}}d_{\vec{\beta}}d_{\vec{\mu}}^2}{d_{\vec{\lambda}'}d_{\vec{\lambda}''}} 
 \frac{f_{\vec{\lambda}'}f_{\vec{\lambda}''}} {f_{\vec{\mu}}}
 \frac{g_{\vec{\lambda}'}g_{\vec{\lambda}''}} {g_{\vec{\mu}}}C^{\vec{\lambda}'}_{\vec{\mu}\vec{\alpha}}C^{\vec{\lambda}''}_{\vec{\mu}\vec{\beta}}.}
\end{equation}

As commented above, to claim that $\mathcal{O}_{\vec{\mu}}$ is a background we must prove that \eqref{entangled} and \eqref{factorized} are equal at $n,N\to\infty$. A rigorous proof of this statement in full generality is hard for the difficulties one encounters when dealing with the Kronecker coefficients. However, I will offer a check, valid for some configurations $\vec{\mu}$, that clearly supports the statement.

The key property which lies under the factorization of \eqref{entangled} into \eqref{factorized} is the factorization of {\it normalized} characters at large $n$\footnote{See \cite{Biane}.}. That is,
\begin{equation}\label{characterfactorization}
    \overline{\chi}_{\mu}(\sigma_1\circ\sigma_2)=\overline{\chi}_{\mu}(\sigma_1)\overline{\chi}_{\mu}(\sigma_2)+o(1/n),
\end{equation}
where the normalized character is defined as
\begin{equation}
    \overline{\chi}_\lambda(\sigma)\equiv \frac{\chi_\lambda(\sigma)}{d_\lambda}.
\end{equation}
The property \eqref{characterfactorization} come from the explicit form of the characters for large $n$ found by Biane \cite{Biane},
\begin{equation}\label{bianecharacter}
    \overline{\chi}_\lambda(\sigma)=C_\sigma(w)\,n^{-|\sigma|/2}+O(n^{-|\sigma|/2-1}),\quad \lambda\vdash n,
\end{equation}
where $|\sigma|$ is the minimal number of transpositions necessary to generate $\sigma$, and $w$ is the limit shape the partition $\lambda$ approaches. The crucial fact in formula \eqref{bianecharacter} is that $C_{\sigma\circ \tau}(w)=C_{\sigma}(w)C_{\tau}(w)$, that is, the function factorizes whenever $\sigma$ and $\tau$ are disjoint permutations. It is also especial that $C_\sigma(w)$ depends only on the limit shape $w$ and not on the particular partition $\lambda$, a fact that I will use later. 

Due to \eqref{characterfactorization}, the Littlewood-Richardson numbers adopt the useful form 
\begin{equation}\label{LRfactorization}
    C^{\lambda}_{\mu\nu}=\frac{1}{n!n_1!}\sum_{\substack{\sigma\in S_{n}\\
    \sigma_1\in S_{n_1}}}\chi_{\lambda}(\sigma\circ\sigma_1)\chi_{\mu}(\sigma)\chi_{\nu}(\sigma_1)\sim \frac{1}{d_\lambda^2}d_{\lambda/\mu} d_{\lambda/\nu},
\end{equation}
where 
\begin{equation}\label{defrel}
  d_{\lambda/\mu}=\frac{1}{n!}\sum_{\sigma\in S_{n}}\chi_{\lambda}(\sigma) \chi_{\mu}(\sigma), \quad \lambda\vdash n+n_1,\, \mu\vdash n,
\end{equation}
 is the number of times that the representation $\Gamma_{\mu}$ is subduced from $\Gamma_{\lambda}$ when the group is restricted from $S_{n+n_1}$ to $S_{n}$. The number $d_{\lambda/\mu}$ also counts the paths that join the partition $\lambda$ and the partition $\mu$ in the Young graph or, equivalently, the number of partially labeled Young diagrams between $\lambda$ and $\mu$. Of course, this number will be 0 if $\mu$ is not subduced by $\lambda$. 
 Usually, the computation of LR numbers, although it can be done combinatorially, is much more complicated than \eqref{LRfactorization}, and involves a precise relation between partitions $\mu$ and $\nu$, besides their individual relation with $\lambda$. However, as seen in \eqref{LRfactorization}, this contribution is subleading at large $n$. \\
 Let us see how  \eqref{entangled} approaches \eqref{factorized} at large $n,N$. The first thing to notice is that the prefactors in front of their respective sums are equal at large $n$, so I will not worry about them in the following. We now have
 \begin{eqnarray}
     &&\sum_{\lambda_i\vdash n+n_1+n_2}\frac{d_{\vec{\alpha}}d_{\vec{\beta}}d_{\vec{\mu}}}{d_{\vec{\lambda}}}\, C^{\vec{\lambda}}_{\vec{\mu}\vec{\alpha}\vec{\beta}}f_{\vec{\lambda}}\,g_{\vec{\lambda}}\approx \sum_{\lambda_i\vdash n+n_1+n_2}\frac{d_{\vec{\lambda}/\vec{\mu}}d_{\vec{\mu}}}{d_{\vec{\lambda}}}\,\frac{d_{\vec{\lambda}/\vec{\alpha}}d_{\vec{\alpha}}}{d_{\vec{\lambda}}}\,\frac{d_{\vec{\lambda}/\vec{\beta}}d_{\vec{\beta}}}{d_{\vec{\lambda}}}\,f_{\vec{\lambda}}\,g_{\vec{\lambda}}\label{line1} \\
     &\approx& \sum_{\substack{
     \tau'_i,\lambda'_i\vdash n+n_1\\
     \tau''_i,\lambda''_i\vdash n+n_2\\
     \vec{\lambda}(\vec{\lambda}',\vec{\lambda}'')}}\frac{d_{\vec{\lambda}''/\vec{\mu}}d_{\vec{\mu}}}{d_{\vec{\lambda}''}}\,\frac{d_{\vec{\lambda}'/\vec{\mu}}d_{\vec{\mu}}}{d_{\vec{\lambda}'}}\,\frac{d_{\vec{\lambda}/\vec{\tau}'}d_{\vec{\tau}'}}{d_{\vec{\lambda}}}\,\frac{d_{\vec{\tau}'/\vec{\alpha}}d_{\vec{\alpha}}}{d_{\vec{\tau}'}}\,\frac{d_{\vec{\lambda}/\vec{\tau}''}d_{\vec{\tau}''}}{d_{\vec{\lambda}}}\,\frac{d_{\vec{\tau}''/\vec{\beta}}d_{\vec{\beta}}}{d_{\vec{\tau}''}}\,f_{\vec{\lambda}}\,g_{\vec{\lambda}}\label{line2}  \\
     &\approx& \sum_{\substack{
     \tau'_i,\lambda'_i\vdash n+n_1\\
     \tau''_i,\lambda''_i\vdash n+n_2\\
     \vec{\lambda}(\vec{\lambda}',\vec{\lambda}'')}}\frac{d_{\vec{\lambda}''/\vec{\mu}}d_{\vec{\mu}}}{d_{\vec{\lambda}''}}\,\frac{d_{\vec{\lambda}'/\vec{\mu}}d_{\vec{\mu}}}{d_{\vec{\lambda}'}}\,\frac{d_{\vec{\lambda}/\vec{\tau}'}d_{\vec{\tau}'}}{d_{\vec{\lambda}}}\,\frac{d_{\vec{\tau}'/\vec{\alpha}}d_{\vec{\alpha}}}{d_{\vec{\tau}'}}\,\frac{d_{\vec{\lambda}/\vec{\tau}''}d_{\vec{\tau}''}}{d_{\vec{\lambda}}}\,\frac{d_{\vec{\tau}''/\vec{\beta}}d_{\vec{\beta}}}{d_{\vec{\tau}''}}\,\frac{f_{\vec{\lambda}'}f_{\vec{\lambda}''}}{f_{\vec{\mu}}}\,\frac{g_{\vec{\lambda}'}g_{\vec{\lambda}''}}{g_{\vec{\mu}}}\label{line3}  \\
     &\approx& \sum_{\substack{
     \tau'_i,\lambda'_i\vdash n+n_1\\
     \tau''_i,\lambda''_i\vdash n+n_2\\
     \vec{\lambda}(\vec{\lambda}',\vec{\lambda}'')}}\frac{d_{\vec{\lambda}''/\vec{\mu}}d_{\vec{\mu}}}{d_{\vec{\lambda}''}}\,\frac{d_{\vec{\lambda}'/\vec{\mu}}d_{\vec{\mu}}}{d_{\vec{\lambda}'}}\,\frac{d_{\vec{\lambda}/\vec{\tau}'}d_{\vec{\tau}'}}{d_{\vec{\lambda}}}\,\frac{d_{\vec{\lambda}'/\vec{\alpha}}d_{\vec{\alpha}}}{d_{\vec{\lambda}'}}\,\frac{d_{\vec{\lambda}/\vec{\tau}''}d_{\vec{\tau}''}}{d_{\vec{\lambda}}}\,\frac{d_{\vec{\lambda}''/\vec{\beta}}d_{\vec{\beta}}}{d_{\vec{\lambda}''}}\,\frac{f_{\vec{\lambda}'}f_{\vec{\lambda}''}}{f_{\vec{\mu}}}\,\frac{g_{\vec{\lambda}'}g_{\vec{\lambda}''}}{g_{\vec{\mu}}}\label{line4}  \\
     &=& \sum_{\substack{
     \lambda'_i\vdash n+n_1\\
     \lambda''_i\vdash n+n_2}}\frac{d_{\vec{\lambda}''/\vec{\mu}}d_{\vec{\mu}}}{d_{\vec{\lambda}''}}\,\frac{d_{\vec{\lambda}'/\vec{\mu}}d_{\vec{\mu}}}{d_{\vec{\lambda}'}}\,\frac{d_{\vec{\lambda}'/\vec{\alpha}}d_{\vec{\alpha}}}{d_{\vec{\lambda}'}}\,\frac{d_{\vec{\lambda}''/\vec{\beta}}d_{\vec{\beta}}}{d_{\vec{\lambda}''}}\,\frac{f_{\vec{\lambda}'}f_{\vec{\lambda}''}}{f_{\vec{\mu}}}\,\frac{g_{\vec{\lambda}'}g_{\vec{\lambda}''}}{g_{\vec{\mu}}}\label{line5}  \\
     &\approx&
     \sum_{\substack{\lambda'_i\vdash n+n_1\\
 \lambda''_i\vdash n+n_2}} \frac{d_{\vec{\alpha}}d_{\vec{\beta}}d_{\vec{\mu}}^2}{d_{\vec{\lambda}'}d_{\vec{\lambda}''}} 
 C^{\vec{\lambda}'}_{\vec{\mu}\vec{\alpha}}C^{\vec{\lambda}''}_{\vec{\mu}\vec{\beta}}\frac{f_{\vec{\lambda}'}f_{\vec{\lambda}''}} {f_{\vec{\mu}}}
 \frac{g_{\vec{\lambda}'}g_{\vec{\lambda}''}} {g_{\vec{\mu}}},\label{line6}
 \end{eqnarray}
 and then,
 \begin{equation}\label{proofappr}
      \langle \mathcal{O}_{\vec{\alpha}} \mathcal{O}_{\vec{\beta}} \mathcal{O}_{\vec{\mu}} \rangle\approx  \frac{ \langle \mathcal{O}_{\vec{\alpha}}\mathcal{O}_{\vec{\mu}}\rangle \langle \mathcal{O}_{\vec{\beta}}\mathcal{O}_{\vec{\mu}}\rangle}{\langle \mathcal{O}_{\vec{\mu}}\rangle}, 
 \end{equation}
 where the equality is reached at the limit $n\to\infty$.\\
 Let us explain the approaches taken in \eqref{line1}-\eqref{line6}. \\
 In \eqref{line1}, I have used \eqref{LRfactorization} to convert the LR-numbers into the dimension and relative dimensions of the irreducible representations associated to the Young graph, the branching graph of the symmetric groups. \\
 In line \eqref{line2}, I have used the chain property\footnote{See \cite{Olshanski}.} of the Young graph
 \begin{equation}
     \frac{d_{\lambda/\alpha}d_{\alpha}}{d_\lambda}=\sum_{\tau}\frac{d_{\lambda/\tau}d_{\tau}}{d_\lambda}\frac{d_{\tau/\alpha}d_{\alpha}}{d_\tau}, \quad \lambda\vdash n\quad \alpha \vdash n'\quad \tau\vdash n'',
 \end{equation}
 valid for all $n\geq n''\geq n'$. I have also made the approximation
 \begin{equation}\label{lambdamu}
   \frac{d_{\lambda/\mu}d_{\mu}}{d_\lambda}\approx \frac{(n_1+n_2)!}{n_1!n_2!}\, \frac{d_{\lambda'/\mu}d_{\mu}}{d_\lambda'}\,\frac{d_{\lambda''/\mu}d_{\mu}}{d_\lambda''}, 
 \end{equation}
 when $\lambda\vdash n+n_1+n_2$ is related to $\lambda'\vdash n+n_1$ and $\lambda''\vdash n+n_2$ as follows. The Young diagram $\lambda'$ is obtained from $\mu$ by adding $n_1$ boxes. We mark the corners of $\mu$ where these boxes are added. We do the same for $\lambda''$ which is obtained from $\mu$ by the addition of $n_2$ boxes. Now, we construct $\lambda$ by adding $n_1+n_2$ boxes to $\mu$ in the indicated corners. Be aware that summing over $\lambda'$ and $\lambda''$ overcounts the sum over $\lambda$ by the factor $\binom{n_1+n_2}{n_1}$, which should divide the sum \eqref{line2}, but it exactly cancels the prefactor in \eqref{lambdamu}. Now, in the approximation  \eqref{lambdamu} it has been taken into account
that
\begin{equation}
   \frac{d_\mu}{d_\lambda}\approx \frac{d_\mu}{d_{\lambda'}}\frac{d_\mu}{d_{\lambda''}}, 
\end{equation}
with the prescription I have given for $\lambda(\lambda',\lambda'')$, as can be easily checked using the hook formula for the dimension of the irreducible representations of the symmetric group
\begin{equation}
    d_\mu=\frac{n!}{\text{Hooks}_\mu}.
\end{equation}
In \eqref{lambdamu}, I have also approached the relative dimensions. Given the Young diagrams $\lambda$ and $\mu$ the relative dimension for large $n$ is well approximated by $(n_1+n_2)!$, and analogously for $\lambda'$ and $\lambda''$. So,
\begin{equation}\label{rellambdamu}
    d_{\lambda/\mu}\approx\frac{(n_1+n_2)!}{n_1!n_2!}d_{\lambda'/\mu}d_{\lambda''/\mu}.
\end{equation}
As said above, the combinatorial factor in \eqref{rellambdamu} exactly cancels the factor that occurs when summing over $\lambda'$ and $\lambda''$ instead of over $\lambda$ in \eqref{line2}.\\
In line \eqref{line3}, I used 
\begin{equation}
    f_\lambda(N)=\prod_{(i,j)\in \lambda}(N+j-i),
\end{equation}
to write
\begin{equation}
    f_{\lambda}=\frac{f_{\lambda'}f_{\lambda''}}{f_{\mu}}.
\end{equation}
For the Kronecker coefficients, the approach is taken under the assumption that the states label by $\vec{\mu}$ are typical states, and so the limit shapes are close to the Pancherel curve, where the highest dimension for representations is reached and Kronecker coefficients are also maximal\cite{Pak}. The word ``typical'' refers to the Pancherel measure
\begin{equation}
    P[\lambda]=\frac{d_\lambda^2}{n!}, \qquad \frac{1}{n!}\sum_{\lambda\vdash n}d_\lambda^2=1,
\end{equation}
 which is naturally associated to the branching graph of the symmetric group. Related to this measure, the probability that we pick the Pancherel curve as the limit shape as $n\to\infty$ is one. 
 
 The value of the Kronecker coefficients for order three tensors when the limit shape is close to the Pancherel curve is
\begin{equation}
    g_{\mu_1\mu_2\mu_3}\approx \sqrt{n!}\exp(-a_p\,n), \qquad \mu_i\vdash n,\quad a_p\geq 0,
\end{equation}
where the equality is reached when all the partitions approach the Pancherel limit shape. This justifies
\begin{equation}\label{apkro}
    g_{\vec{\lambda}}\approx\frac{g_{\vec{\lambda}'}g_{\vec{\lambda}''}}{g_{\vec{\mu}}},
\end{equation}
used in \eqref{line3}. As commented above, the approach of Kronecker coefficients \eqref{apkro} is modest since it is restricted to limit shapes close to the Pancherel curve. I claim that the factorization of the three-point function \eqref{proofappr} occurs for {\it any} limit shape, so the definite proof will have to involve a generalization of \eqref{apkro}, valid for generic limit shapes. 

In \eqref{line4}, the only changes I introduce are
\begin{equation}
    \frac{d_{\vec{\tau}'/\vec{\alpha}}}{d_{\vec{\tau}'}}\longrightarrow \frac{d_{\vec{\lambda}'/\vec{\alpha}}}{d_{\vec{\lambda}'}} \quad \text{and} \quad \frac{d_{\vec{\tau}''/\vec{\beta}}}{d_{\vec{\tau}''}}\longrightarrow \frac{d_{\vec{\lambda}''/\vec{\beta}}}{d_{\vec{\lambda}''}}.
\end{equation}
Be aware that, from \eqref{defrel}, we see that
\begin{equation}\label{proofrels}
    \small{\frac{d_{\tau'/\alpha}}{d_{\tau'}}=\frac{1}{n_1!}\sum_{\sigma\in S_{n_1}}\overline{\chi}_{\tau'}(\sigma)\chi_\alpha(\sigma)\approx \frac{1}{n_1!}\sum_{\sigma\in S_{n_1}} C_\sigma(w)n^{-|\sigma|/2}\chi_\alpha(\sigma)\approx \frac{1}{n_1!}\sum_{\sigma\in S_{n_1}}\overline{\chi}_{\lambda'}(\sigma)\chi_\alpha(\sigma)=\frac{d_{\lambda'/\alpha}}{d_{\lambda'}}.}
\end{equation}
The key point in \eqref{proofrels} is that, both $\tau'$ and $\lambda'$ differ in $n_1$ boxes from $\mu$, and they have the same limit shape $w$. \\
Finally, in \eqref{line5}, I have used the stochastic property of the relative dimensions \cite{Olshanski}, by means of which,
\begin{equation}
    \sum_{\tau'}\frac{d_{\lambda/\tau'}d_{\tau'}}{d_{\lambda}}=1.
\end{equation}

Note that the factorization of the three-point function \eqref{proofappr} is a very non-trivial statement. The factorization properties of characters at large $n$ play a crucial role in this approach. The product of Schur invariants involves LR numbers, whose approximation \eqref{LRfactorization} at large $n$ is at the core of the proof. As I said, Schur invariants are perhaps not the only ones that are entitle to partner backgrounds but I do not find it easy to think of other invariants who fulfil \eqref{proofappr}, and can be proven so.

\subsection{Path integral ansatz}\label{PIA}

Although it is not clear, and certainly not proven in this article, that Schur invariants are the only large states which can be traded as backgrounds, we find it convenient to restrict ourselves to this subspace of invariants. The reason is  two-fold: on the one hand, I have just proven in \eqref{proofappr} that they behave appropriately at large $n$, and on the other, they are easier to operate with since we can construct them explicitly. 

In the following I am going to be loyal to two ideas:
\begin{enumerate}
    \item I will take seriously the idea that tensor models encode  quantum gravity. This is a reasonable assumption given the success of tensor models describing discretized quantum gravity via triangulations.
    \item I will associate a background to a Schur operator $\mathcal{O}_{\vec{\mu}}$. This is analogous to the usual association trace invariant $\leftrightarrow$ PL-manifold. However, as opposed to the triangulation scheme, it is not obvious  how to make the association Schur operator $\leftrightarrow$ PL-manifold {\it a priori}.  
\end{enumerate}
 
 As said above, the interacting terms I will be considering in the action are Schur invariants. Thus, the action will be 
 \begin{equation}\label{actionschurs}
     S[T,\overline{T}]= T\cdot\overline{T}+\frac{1}{N^{d-1}}\sum_{\vec{\mu}}\lambda_{\vec{\mu}}\mathcal{O}_{\vec{\mu}}.
 \end{equation}
Therefore, the partition function of the tensor model I am considering is \eqref{pftm} with \eqref{actionschurs}. 
In the proposal I am making, the partition function of the tensor model should be equated to the partition function of gravity, so one would like to schematically write
\begin{equation}\label{ZequalZ}
    Z[\lambda]=Z_g[\kappa]=\int dg\,\exp\big(-S_{\kappa}[g]\big),
\end{equation}
where $\kappa$ is a label for the higher derivative terms of the gravity action.  
In order to make some sense from \eqref{ZequalZ}, let us examine the region near a solution of the gravity equations, that is, near a background. In view of the correspondence between Schur invariants and backgrounds, and with a slight abuse of notation, let us refer to the background as $\vec{\mu}$, when associated with the tensor invariant $\mathcal{O}_{\vec{\mu}}$. Accordingly, I will write $S[\vec{\mu}]$ for the on-shell gravity action on the background $\vec{\mu}$. Near this background, the gravity path integral can be well approximated as
\begin{equation}\label{sadlepoint}
  Z_g[\kappa]=e^{-S_{\kappa}[\vec{\mu}]}.  
\end{equation}
On the tensor side of the equality \eqref{ZequalZ}, we can write
\begin{equation}\label{Zexp}
   Z[\lambda]=\big\langle \exp \big(\sum_{\vec{\mu}}\lambda_{\vec{\mu}}\mathcal{O}_{\vec{\mu}}\big)\big\rangle_0,
\end{equation}
where the subscript 0 reminds us that the average is Gaussian. Note that in \eqref{Zexp} there is no Schur invariant chosen, the sum is over all of them. In order to match the saddle point approach of \eqref{sadlepoint}  with its tensor counterpart we have to impose a certain ``projection'' of $Z[\lambda]$ onto $\mathcal{O}_{\vec{\mu}}$. I will write
\begin{equation}\label{eqPF}
    P_{\vec{\mu}} \big(Z[\lambda]\big) = e^{-S_{\kappa}[\vec{\mu}]}. 
    \end{equation}
For the purpose of finding such projection, let us remember that since $N$ must be large to make contact with classical gravity and because tensor models are asymptotically free \cite{G1}, the couplings $\lambda_{\vec{\mu}}(N)$ must be small and a Taylor expansion on them is expected to be accurate with a few terms. So, let us Taylor expand \eqref{Zexp}. We will have
\begin{equation}
 Z[\lambda]=\sum_{n\geq0}\frac{1}{n!}\big\langle \Big(\sum_{\vec{\nu}}\lambda_{\vec{\nu}}\mathcal{O}_{\vec{\nu}}\Big)^n\big\rangle_0.
\end{equation}
The first terms of this expansion read
\begin{equation}\label{Zexpandterms}
 Z[\lambda]=1+\sum_{\vec{\nu}}\lambda_{\vec{\nu}}\big\langle\mathcal{O}_{\vec{\nu}}\big\rangle_0+\frac{1}{2} \sum_{\vec{\nu}_1,\vec{\nu}_2}\lambda_{\vec{\nu}_1}\lambda_{\vec{\nu}_2}\big\langle\mathcal{O}_{\vec{\nu}_1}\mathcal{O}_{\vec{\nu}_2}\big\rangle_0 +\dots
\end{equation}
    The projection of $Z[\lambda]$ onto $\mathcal{O}_{\vec{\mu}}$ is straightforward at the sight of \eqref{Zexpandterms}.  The product of two Schur operators with $2n_1$ and $2n_2$ tensors is again a Schur operator 
    \begin{equation}
        \mathcal{O}_{\vec{\nu}_1}\mathcal{O}_{\vec{\nu}_2}=\sum_{\vec{\mu}} a_{\vec{\nu}_1\vec{\nu}_2}^{\vec{\mu}}\mathcal{O}_{\vec{\mu}},
    \end{equation}
    where $\mathcal{O}_{\vec{\mu}}$ is made of $2n=2n_1+2n_2$ tensors, and $a_{\vec{\nu}_1\vec{\nu}_2}^{\vec{\mu}}$ is proportional to the product of Littlewood-Richardson numbers $C_{\vec{\nu}_1\vec{\nu}_2}^{\vec{\mu}}$, see the appendix \ref{prodschur}. Higher order terms in the expansion are similar, involving coefficients $C_{\vec{\nu}_1\vec{\nu}_2\vec{\nu}_3}^{\vec{\mu}},C_{\vec{\nu}_1\vec{\nu}_2\vec{\nu}_3\vec{\nu}_4}^{\vec{\mu}}$, and so on. Thus, the natural projection of the expansion \eqref{Zexpandterms} onto $\mathcal{O}_{\vec{\mu}}$ reads
    \begin{eqnarray}\label{expansionZtensor}
      P_{\vec{\mu}} \big(Z[\lambda]\big)&=&1+ \lambda_{\vec{\mu}}\big\langle\mathcal{O}_{\vec{\mu}}\big\rangle_0 +
      \sum_{\vec{\nu}_1,\vec{\nu}_2}\lambda_{\vec{\nu}_1}\lambda_{\vec{\nu}_2}a_{\vec{\nu}_1\vec{\nu}_2}^{\vec{\mu}}\big\langle\mathcal{O}_{\vec{\mu}}\big\rangle_0+\dots\nonumber \\
      &=& 1+\big\langle\mathcal{O}_{\vec{\mu}}\big\rangle_0 \Big(\lambda_{\vec{\mu}}+\sum_{\vec{\nu}_1,\vec{\nu}_2}\lambda_{\vec{\nu}_1}\lambda_{\vec{\nu}_2}a_{\vec{\nu}_1\vec{\nu}_2}^{\vec{\mu}}+\dots \Big).
    \end{eqnarray}
  Similarly, we can Taylor expand \eqref{sadlepoint} with respect to the coefficients of the higher derivative terms. For instance, in four dimensions, the most general  quadratic, covariant, parity-invariant, metric-compatible and torsion-free action is \cite{BGKM} 
  \begin{equation}\label{IDG}
      S_{F_1,F_2,F_3}[g]=\int d^4x\,\frac{1}{2}\sqrt{g}\Big(M_p^2R+RF_1(\square)R+R_{ab}F_2(\square)R^{ab}+R_{abcd}F_3(\square)R^{abcd}\Big),
  \end{equation}
    with
    \begin{equation}
        F_i(\square)=\sum_{n=0}^{\infty}f_{i_n}\frac{\square^n}{M^{2n}}, \quad \square=g^{ab}\nabla_a\nabla_b,
    \end{equation}
    with $M\leq M_p$ a certain mass scale that cannot be too small so that Einstein action is still accurate for the current observations. The proposed prescription for the gravity side is  to first evaluate the action at the background $\mu$, and then perform the Taylor expansion of $e^{-S_{\kappa}[\vec{\mu}]}$ with respect to the coefficients $\frac{f_{i_n}}{M^{2n}}$.  The result can be reorganised as
    \begin{equation}\label{GTE}
        e^{-S_{\kappa}[\vec{\mu}]}=1+\gamma_1 S_1[\vec{\mu}]+\gamma_2 S_2[\vec{\mu}]+\dots,
    \end{equation}
where $S_1[\vec{\mu}]$ is the HE action evaluated at $\vec{\mu}$, $S_2[\vec{\mu})$ will be the action of the terms quadratic in curvatures evaluated at $\vec{\mu}$, and so on. The coefficients $\gamma$ in \eqref{GTE} depend on products of the functions $f_i$ and on the scale. The dependence on the scale is
\begin{equation}
 \gamma_1=-\frac{M_p^2}{2}, \quad   \gamma_i\sim M^{2-2i},\,\, i\geq 2.
\end{equation}
Now, we equate \eqref{expansionZtensor} with \eqref{GTE}. At leading order, we have
\begin{equation}\label{eqfo}
  \boxed{\big\langle\mathcal{O}_{\vec{\mu}}\big\rangle_0 =S_{HE}[\vec{\mu}]}\,.    
\end{equation}

\section{Ward identities of tensor models and locality}\label{Wid}
The Ward identities arise in tensor models associated with the change of integration variables in the partition function. That is with the change $T\rightarrow T+\delta T$ and $\overline{T}\rightarrow \overline{T}+\delta \overline{T}$ given by
\begin{eqnarray}
    T_{i_1,\dots,i_d} &\longrightarrow& T_{i_1,\dots,i_d}+\frac{\delta \mathcal{O}}{\delta T^{i_1,\dots,i_d}},\nonumber \\
    \overline{T}_{i_1,\dots,i_d} &\longrightarrow& \overline{T}_{i_1,\dots,i_d}+\frac{\delta \mathcal{O}}{\delta \overline{T}_{i_1,\dots,i_d}},
\end{eqnarray}
where $\mathcal{O}$ is a gauge invariant operator. Using these transformations, it is found that the symmetries of the action translate into a tower of identities among averages, see \cite{ito1,ito2,ito3}. For an action given by \eqref{actionschurs}, these identities may be written as\footnote{See \cite{ito3}, equation (2.2).}
\begin{equation}\label{wardidentities}
N^{d-1}|\mathcal{O}|\,\big\langle \mathcal{O}\big\rangle= \sum_{\mathcal{O}'}\lambda_{\mathcal{O}'}\,\big\langle \{\mathcal{O},\mathcal{O}'\}\big\rangle +\big\langle \Delta \mathcal{O}\big\rangle,
\end{equation}
where $|\mathcal{O}|=n$ for an invariant made of $2n$ tensors, the sum is over all invariants $\mathcal{O}'$ present in the action (Schur invariants) and the averages are taken with the full action. I have denoted
\begin{align}
\label{CUT}
\Delta \mathcal{O}&= \frac{\delta^2 \mathcal{O}[T]}{\delta T_{i_1\dots i_d} \delta \overline{T}^{i_1\dots i_d}},  \\
\{\mathcal{O},\mathcal{O}'\}&= \frac{\delta \mathcal{O}[T]}{\delta T_{i_1\dots i_d}}\, \frac{\delta \mathcal{O}'[T]}{\delta\overline{T}^{i_1\dots i_d}}\label{JOIN},
\end{align}
which are the {\it cut} and {\it join} operators, respectively, defined in \cite{ito2}. Similar operators, with different names, also appear in the literature. For instance, in \cite{R4}, these operators are called $\omega$ and $\Omega$, respectively.\\
For Gaussian averages, \eqref{wardidentities} turns into
\begin{equation}\label{Wardgaussian}
    \big\langle \mathcal{O}\big\rangle_0= \frac{1}{nN^{d-1}}\big\langle \Delta \mathcal{O}\big\rangle_0.
\end{equation}
We now apply \eqref{Wardgaussian} to \eqref{eqfo}, and obtain
\begin{equation}\label{localterm}
\boxed{\frac{1}{nN^{d-1}}\big\langle \Delta \mathcal{O}_{\vec{\mu}}\big\rangle_0=S_{HE}[\vec{\mu}].}
 \end{equation}
In the next subsection, I am going to show that emergence of spacetime and a notion of locality are already present in \eqref{localterm}.

\subsection{The spacetime grid}

It is proven in appendix \ref{CAODS}  that
\begin{equation}\label{cutepansion}
  \Delta \mathcal{O}_{\vec{\mu}}=\sum_{\vec{\mu}'\nearrow \vec{\mu}}C(\vec{\mu},\vec{\mu}')\mathcal{O}_{\vec{\mu}'},\quad \text{with }\quad C(\vec{\mu},\vec{\mu}')=n^3\frac{g_{\vec{\mu}}}{g_{\vec{\mu}'\uparrow \vec{\mu}}}\sum_{\vec{\mu}'\nearrow \vec{\mu}}\frac{\text{Dim}_{\vec{\mu}}(N)}{\text{Dim}_{\vec{\mu}'}(N)},  
\end{equation}
where $\vec{\mu}'\nearrow \vec{\mu}$ indicates  the collection of $d$ partitions of $n-1$ elements that appear as we delete one of the corners of each of the $d$ partitions of $n$ elements $\vec{\mu}$. The are 
as many $\vec{\mu}'$'s in the expansion \eqref{cutepansion} as the product of the number of corners of the $d$ partitions in $\vec{\mu}$. Since each partition $\mu_i$ has at most $N$ parts, its number of corners is at most equal to $N$. Let us associate to each corner $c_j(\mu_i)$ the number of the row it appears at. For instance,
\begin{equation}
\young(\hfil \hfil 1,\hfil \hfil ,\hfil  3,\hfil ,5).
\end{equation}
With this association every $\vec{\mu}'$  of $\vec{\mu}'\nearrow \vec{\mu}$ is mapped to
a $d$-tuple made by choosing the value of the deleted corner in each diagram $\mu_i$. 
Let us see an example. Consider, for $d=3$ and $n=5$,
\begin{equation}
    \vec{\mu}=(\,\, \young(\hfil\hfil 1,\hfil 2), \young(\hfil\hfil 1,\hfil ,3), \young(\hfil 1,\hfil ,\hfil ,4)),
\end{equation}
where we have already labeled the corners. Then, one of the subduced $\vec{\mu}'$'s is
\begin{equation}
   \vec{\mu}'= (\,\, \young(\hfil\hfil \hfil,\hfil ), \young(\hfil\hfil ,\hfil ,\hfil), \young(\hfil \hfil ,\hfil ,\hfil )) \Longleftrightarrow (2,1,4),
\end{equation}
and another 
\begin{equation}
   \vec{\mu}'= \Bigg(\,\, \young(\hfil \hfil,\hfil \hfil ), \young(\hfil\hfil ,\hfil ,\hfil), \young(\hfil ,\hfil ,\hfil ,\hfil )\,\bigg) \Longleftrightarrow (1,1,1),
\end{equation}

Let us consider a $N^d$ grid parameterized by $\vec{r}$ with $ r_i=1,\dots,N$.  With the above prescription, each $\vec{\mu}'$ can be mapped to a point in the grid
\begin{equation}\label{gridmap}
   \vec{\mu}' \nearrow \vec{\mu}\longmapsto \vec{r}.
\end{equation}
Since the number of corners of each $\mu_i$ is always less or equal than $N$, in general not all the points in the grid get occupied, but the map is injective. Thus, given a label $\vec{\mu}$, there is no point in the grid who is mapped to two different $\vec{\mu}'$'s.  Now, with this map any sum $\sum_{\vec{\mu}'\nearrow \vec{\mu}}$, and specifically \eqref{cutepansion}, can be understood as a sum  over the grid. For that, the function $C(\vec{\mu},\vec{\mu}')$ must be interpolated so that it takes values all over the grid. On the RHS of equation \eqref{localterm} we have a sum over the whole volume of spacetime which, by splitting each coordinate in $N$ parts, can be discretized and turned into the sum
\begin{equation}
    \int_V d^dx R\sqrt{g}\longrightarrow \sum_{\vec{r}\,\in\,\text{grid}}R(\vec{r})
\end{equation}
The equation \eqref{localterm} turns into an equality between two sums over the grid. The equation is automatically fulfilled if we identify the summands on both sides of the equation. In this case, we obtain
\begin{equation}\label{eqatthepoint}
    \frac{1}{nN^{d-1}} C(\vec{r})\,\big\langle \mathcal{O}_{\vec{r}}\big\rangle_0=R(\vec{r})
\end{equation}
where, on the LHS, we have already performed the map $\vec{\mu}'\to\vec{r}$. Equation \eqref{eqatthepoint} provides a prescription for the emergence of a background in tensor models.

\subsection{Locality}
One of the nice features of the picture proposed in equation \eqref{eqatthepoint} is the emergence of a sense of locality in tensor models. Nearby points in the grid correspond to nearby corners in the Young diagrams $\vec{\mu}$. 
The quantity on the LHS of \eqref{eqatthepoint} varies from one point to the grid to another according to the ``corner distance'' in the Young diagrams which, for adjacent corners is minimal. It will be an infinitesimal when we take the continuum limit. If we want to reconstruct a background from a tensor model, this fact translates, via \eqref{eqatthepoint}, into a smooth curvature and thus a regular manifold. In short, the main message of \eqref{eqatthepoint} is the fact that the sense of locality in tensor models is related to nearby corners in the Young diagrams which label Schur invariants. A similar idea of locality  was already suggested in the context of holography in \cite{RdMK}, and recently developed in \cite{R3}. 

\subsection{ADM mass}\label{BH}
For tensor models with matter, we would have a variety of backgrounds. Regular manifolds, with non-singular curvature, will relate to balanced Young diagrams. Remember that balanced diagrams, or limit shapes, are the ones which label regular irreducible representations of $S_\infty$ \cite{Th}. Those diagrams can be constructed inductively by adding boxes and rescaling, in a way that in the limit $n\to \infty$ we obtain a monotonous function enclosing a non-zero area with the axis. Thus, $\vec{\mu}\longrightarrow (f_1,\dots,f_d)$. The non-zero area condition, together with the fact that no Young diagram exceeds $N$ rows lead to the conclusion that $n\sim N^2$, which is also the relation found in holography  for the states that relate to new geometries. 
Thus, in the classical limit, with $N,n\to \infty$, the relevant quantity will be
\begin{equation}
    \rho=\frac{n}{N^2}\in \mathbb{R}_+,
\end{equation}
and we expect all non-zero non-divergent classical quantities to depend on $n$ and $N$ only through $\rho$. 

Let us consider asymptotically flat spacetimes. It is known that those spaces admit an ADM mass, that is, a mass-like charge associated to the whole spacetime. It is natural to wonder what this charge is in the tensorial description and if we can read it off from the labels of Schur operators.

The positive number $\rho$ is an intrinsic quantity of Schur operators and ranges from 0 to $\infty$. It is related to the energy of the state produced as the Schur operator acts on the vacuum state. Indeed, the energy of the Schur states must grow linearly with $n$, as it does for the harmonic oscillator\footnote{One can take the analogy with the harmonic oscillator further. The tensor model can be understood as a collection of $N^d$ identical harmonic oscillators, one for each component of the tensor. Then each component acts as the creation operator and its conjugate as the annihilation operator when acting on the vacuum state. With this picture, an invariant made of $n$ tensors  acting on the vacuum produces a sum of excited states, all with excitation number $d\,n$.}, recovering the vacuum for $n=0$. These considerations make $\rho$ a good candidate
for the ADM mass of the associated geometry. Thus, I propose the ADM mass to be given by the quantity
\begin{equation}\label{ADMansatz}
    M_{ADM}=C \rho,
\end{equation}
where $C$ does not depend on either $n$ nor $N$. Note that for $n=0$ we have flat spacetime. \\
Using \eqref{ADMansatz} and suitable configurations $\vec{\mu}$ compatible with black hole geometries (based on symmetry arguments, probably), it would be interesting to compute the Bekenstein-Hawking entropy and compare it with the direct counting of (tensor) microstates. This would provide a valuable check of this proposal. The task does not seem easy, however, since it involves counting the number of invariants $\mathcal{O}_{\vec{\mu},ij}$ compatible with the symmetries what, in the end, translates into counting sums of Kronecker coefficients. I will consider it in a future work.

\section{Summary and outlook}
In this paper I have provided a picture of emergent spacetime by making a connection between tensor models at the large $N$ and $n$ limit and classical gravity. In contrast with the usual correspondence between permutation invariants and triangulations, in this proposal Schur invariants take a prominent role and are mapped to backgrounds. At the heart of this proposal is the reorganization of the space of invariants into  a basis driven by representation theory, the restricted Schur basis, which is found suitable for the connection with gravity at the large $N$ and $n$ limit.  

Without a background it is hard to make any sense of locality. Thus, it is always challenging for a background independent theory of gravity to incorporate local theories which, at the end of the day, are the ones that describe most of the physics phenomena we are able to test. In my proposal, by the use of Ward identities, I am able to offer a natural sense of locality in the tensor invariant which encode backgrounds: two points in the grid are close if the corners of the Young diagrams  they are mapped to are close. This way, the corner distance in the Young diagrams that label the Schur invariants in the tensor model translates into physical distance in gravity.

There are a number of lines I find interesting to explore in future works. The correspondence between tensorial and gravitational quantities  is still incomplete. More examples will provide a more detailed picture of the interrelation between both theories. For instance, it would be very interesting to reproduce the BH entropy in the tensor picture, by counting the tensor states compatible with the Schwarzschild geometry. Besides, the tensor theory (and the full gravity theory) are not determined in this paper since the couplings of the different interaction terms in the action are not fixed. They could be fixed by a sensible renormalization flow equation \cite{WiP}. Interestingly, a Wetterich type of equation has already been proposed in tensor models \cite{astrid}, where the sector corresponding to gravity sits at a fix point of the flow. This way, universality is expected to wipe off all the spurious details of discretizations.

\vspace{0.2cm}
\noindent
{\bf Acknowledgment}

\noindent
I would like to thank Robert de Mello Koch for his valuable feedback, encouragement and support regarding both conceptual and technical aspects of the paper. This work has been supported by Universidad de Zaragoza.

\appendix

\section{Restricted Schur basis and Schur invariants}
\subsection{Schur invariants}\label{SO}
Since the observables are build on $n$ indistinguishable copies of $T$ and $\overline{T}$, it is straightforward to see that operators in \eqref{spanset} enjoy the symmetry of shuffling the copies of $T$ and $\overline{T}$ independently, so one must consider
\begin{equation}\label{sym}
    \mathcal{O}_{\vec{\alpha}}\sim\mathcal{O}_{\sigma\cdot\vec{\alpha}\cdot\tau}, \quad \sigma, \tau\in S_n.
\end{equation}
A generic observable is a linear combination of generators $\mathcal{O}_{\vec{\alpha}}$, so it can be written as
\begin{equation}
    \mathcal{O}_f=\sum_{\vec{\alpha}}f(\vec{\alpha})\mathcal{O}_{\vec{\alpha}},
\end{equation}
where $f:S_n^{ d}\longrightarrow \mathbb{C}$.\\
Owing to the symmetry \eqref{sym}, a basis of observables is given by a basis of complex functions $f(\vec{\alpha})$ with the property
\begin{equation}\label{symf}
    f(\sigma\cdot\vec{\alpha}\cdot\tau)=f(\vec{\alpha}),\quad \sigma,\tau\in S_n,
\end{equation}
which will be called from now on {\it double coset invariant} (DCI) functions.\\
Note that the problem is analogous to finding class functions of symmetric group, functions with the property $\chi(\sigma \alpha\sigma^{-1})=\chi(\alpha)$. These latter functions are the well-known characters of the symmetric group. The task, in order to obtain a basis of observables, is therefore to find a basis of double coset functions.\\
The usual convolution algebra of functions of the symmetric group can be extended to the double coset invariant functions as
\begin{equation}\label{algebraf}
h(\vec{\alpha})=f* g \,(\vec{\alpha})=\sum_{\vec{\beta}\in S_n^{ d}} f(\vec{\beta})g(\vec{\beta}^{-1}\cdot\vec{\alpha}).
\end{equation}
It is easy to see that $h(\vec{\sigma\alpha\tau})=h(\vec{\alpha})$, so \eqref{algebraf}  defines an algebra of double coset functions. This algebra is non commutative, but it is associative and it has unit function
\begin{equation}
    f*\delta_{DCI} \,(\vec{\alpha})= \delta_{DCI} *f\, (\vec{\alpha})=f(\vec{\alpha}). 
\end{equation}
The unit element $\delta_{DCI}$ is constructed by means of the delta function of the symmetric group $\delta(\sigma)$ which is 0 unless $\sigma$ is the identity, in which case it is 1. Thus,
\begin{equation}
    \delta_{DCI}(\vec{\alpha})=\sum_{\sigma\in S_n}\delta(\vec{\alpha}\cdot \sigma),
\end{equation}
which is double coset invariant, and it is 0 unless $\vec{\alpha}=(\tau,\dots,\tau)$, for any $\tau\in S_n$. Using the identity
\begin{equation}
    \delta(\sigma)=\frac{1}{n!}\sum_{\mu\vdash n} d_{\mu}\chi_{\mu}(\sigma),  
    \end{equation}
the unit element $\delta_{DCI}(\vec{\alpha})$ can also be put in terms of characters of the symmetric group as
\begin{equation}\label{deltadc}
    \delta_{DCI}(\vec{\alpha})=\frac{1}{n!^d}\sum_{\vec{\mu}}\sum_{\sigma\in S_n}d_{\vec{\mu}}\chi_{\vec{\mu}}(\vec{\alpha}\cdot \sigma).
\end{equation}
The expansion of the unit function \eqref{deltadc} indicates that the function defined as

\begin{equation}\label{projectorsmu}
    \mathcal{P}_{\vec{\mu}}(\vec{\alpha})=\frac{1}{n!^d}\sum_{\sigma\in S_n}d_{\vec{\mu}}\chi_{\vec{\mu}}(\vec{\alpha}\cdot \sigma)
\end{equation}
projects onto the subspace of operators labeled by $\vec{\mu}$. \\
It is easy to see that the Schur invariants $\mathcal{O}_{\vec{\mu}}$ are driven by projectors. That is,
\begin{equation}
  \mathcal{O}_{\vec{\mu}}= \sum_{i=1}^{g_{\vec{\mu}}} \mathcal{O}_{\vec{\mu};ii}= \sum_{\vec{\alpha}} \mathcal{P}_{\vec{\mu}}(\vec{\alpha})\mathcal{O}_{\vec{\alpha}}.
\end{equation}
Schur invariants form a distinguished sector of the restricted Schur basis, and are going to play a predominant role in this paper as partners of backgrounds.

\subsection{Restricted Schur basis and correlators}\label{RSB}
With the Schur basis of DCI functions
\begin{equation}\label{basisF}
    I=\{F_{\vec{\mu};ij}(\vec{\alpha)}|\quad \mu_i\vdash n,\quad l(\mu_i)\leq N, \quad i,j=1,\dots, g_{\vec{\mu}}\},
\end{equation}
the convolution algebra \eqref{algebraf} can be written as
\begin{equation}\label{algebraF}
  F_{\vec{\mu};ij}*F_{\vec{\nu};kl}(\vec{\alpha})=\delta_{\vec{\mu}\vec{\nu}}\delta_{jk}F_{\vec{\mu};il}(\vec{\alpha}).  
\end{equation}
Since DCI fucntions act naturally on trace operators to produce operator invariants, the algebra \eqref{algebraF} induces an algebra in the space of invariant operators. 
The algebra \eqref{algebraF} is also compatible with the involution
\begin{equation}\label{involution}
    \overline{F}_{\vec{\mu};ij}(\vec{\alpha})=F_{\vec{\mu};ji}(\vec{\alpha}^{-1}).
\end{equation}
Let us define the matrix $\big(\mathcal{M}_{\vec{\mu}}(\vec{\alpha})\big)_{ij}$ of size $g_{\vec{\mu}}\times g_{\vec{\mu}}$ which contains the function $F_{\vec{\mu};ij}(\vec{\alpha})$ at the site $(ij)$. Note that a unitary transformation with $U(g_{\vec{\mu}})$ acting on $\mathcal{M}_{\vec{\mu}}(\vec{\alpha})$ as $U(g_{\vec{\mu}})_{ij}\big(\mathcal{M}_{\vec{\mu}}(\vec{\alpha})\big)_{jk}U^{-1}(g_{\vec{\mu}})_{kl}$ does not alter the convolution structure \eqref{algebraF}.
In view of \eqref{involution} it is easy to see that  $\mathcal{M}_{\vec{\mu}}(\vec{1})$ is self-adjoint. This means that by rearranging the basis with a unitary transformation we can make $\mathcal{M}_{\vec{\mu}}(\vec{1})$ diagonal, that is,   $F_{\vec{\mu};ji}(\vec{1})\propto \delta_{ij}$. Moreover, we can always choose a convenient {\it normalization} so that
\begin{equation}\label{F1}
    F_{\vec{\mu};ji}(\vec{1})=\delta_{ij}.
\end{equation}
Thus, our basis of DCI functions will fulfill \eqref{F1}. Now, a basis of invariant operators $\{\mathcal{O}_{\vec{\mu},ij}\}$ is obtained by acting with each element of \eqref{basisF} on trace invariants as
\begin{equation}\label{Schurbasis}
    \mathcal{O}_{\vec{\mu},ij}=\sum_{\vec{\alpha}}F_{\vec{\mu};ji}(\vec{\alpha})\mathcal{O}_{\vec{\alpha}}.
\end{equation}

Now, let us compute the expectation value of the elements of the basis \eqref{basisF}. Using \eqref{corretr}, the Gaussian averages of the elements of the restricted Schur basis read
\begin{eqnarray}
\langle \mathcal{O}_{\vec{\mu};ij}\rangle_0 &=& \sum_{\vec{\alpha}}F_{\vec{\mu};ij}(\vec{\alpha}) \langle \mathcal{O}_{\vec{\alpha}}\rangle_0= \frac{1}{N^{n(d-1)}}\sum_{\vec{\alpha}}\sum_{\sigma\in S_n} F_{\vec{\mu};ij}(\vec{\alpha})N^{C(\vec{\alpha}\cdot \sigma)}\nonumber \\ &=&\frac{1}{N^{n(d-1)}}\sum_{\vec{\alpha},\vec{\nu}}\sum_{\sigma\in S_n} F_{\vec{\mu};ij}(\vec{\alpha})\chi_{\vec{\nu}}(\vec{\alpha}\cdot \sigma)\text{Dim}_{\vec{\nu}}(N)\nonumber \\
&=&\frac{n!^d}{N^{n(d-1)}}\sum_{\vec{\nu},\vec{\alpha}}F_{\vec{\mu};ij}(\vec{\alpha})\bigg[\frac{1}{n!^d}\sum_{\sigma} d_{\vec{\nu}}\chi_{\vec{\nu}}(\vec{\alpha}\cdot \sigma)\bigg]\frac{\text{Dim}_{\vec{\nu}}(N)}{d_{\vec{\nu}}}\nonumber \\
&=&\frac{n!^d}{N^{n(d-1)}}\sum_{\vec{\nu},\vec{\alpha}}F_{\vec{\mu};ij}(\vec{\alpha})\mathcal{P}_{\vec{\nu}}(\vec{\alpha})\frac{\text{Dim}_{\vec{\nu}}(N)}{d_{\vec{\nu}}}\nonumber \\
&=&\frac{n!^d}{N^{n(d-1)}} F_{\vec{\mu};ij}(\vec{1})\frac{\text{Dim}_{\vec{\mu}}(N)}{d_{\vec{\mu}}}\nonumber \\
&=&\delta_{ij}\frac{n!^d}{N^{n(d-1)}}\frac{\text{Dim}_{\vec{\mu}}(N)}{d_{\vec{\mu}}}.
\end{eqnarray}
The Gaussian average of Schur invariants $\mathcal{O}_{\vec{\mu}}$ are now straightforwardly computed. They are
\begin{equation}
    \langle \mathcal{O}_{\vec{\mu}}\rangle_0 =\sum_{i=1}^{g_{\vec{\mu}}}\langle \mathcal{O}_{\vec{\mu};ii}\rangle_0= \frac{n!^d}{N^{n(d-1)}} \frac{\text{Dim}_{\vec{\mu}}(N)}{d_{\vec{\mu}}}g_{\vec{\mu}}.
\end{equation}

\section{Product of two Schur operators}\label{prodschur}
The Littlewood-Richardson numbers can be computed as
\begin{equation}
C_{\mu\nu}^{\lambda}=\frac{1}{n!m!}\sum_{\sigma\in S_n}\sum_{\tau\in S_m}\chi_{\mu}(\sigma)\chi_{\nu}(\tau)\chi_{\lambda}(\sigma\circ\tau),\quad \mu\vdash m, \,\,\, \nu\vdash n, \,\, \lambda\vdash n+m.
\end{equation}
Now, the product of two Schur operators, one with $2n$ tensors and the other with $2m$, is
\begin{eqnarray}\label{productofSchurs}
    \mathcal{O}_{\vec{\mu}}\,\mathcal{O}_{\vec{\nu}}&=&\frac{1}{(n!m!)^{d-1}}\sum_{\vec{\alpha}\in S_n^d}\sum_{\vec{\beta}\in S_m^d}d_{\vec{\mu}}d_{\vec{\nu}}\chi_{\vec{\mu}}(\vec{\alpha})\chi_{\vec{\nu}}(\vec{\beta})\mathcal{O}_{\vec{\alpha}}\,\mathcal{O}_{\vec{\beta}}\nonumber \\
    &=&\frac{d_{\vec{\mu}}d_{\vec{\nu}}}{(n!m!)^{d-1}}\sum_{\vec{\alpha}\in S_n^d}\sum_{\vec{\beta}\in S_m^d}\chi_{\vec{\mu}}(\vec{\alpha})\chi_{\vec{\nu}}(\vec{\beta})\mathcal{O}_{\vec{\alpha}\circ \vec{\beta}}\nonumber \\
    &=&\frac{d_{\vec{\mu}}d_{\vec{\nu}}}{(n!m!)^{d-1}}\sum_{\vec{\alpha}\in S_n^d}\sum_{\vec{\beta}\in S_m^d}\chi_{\vec{\mu}}(\vec{\alpha})\chi_{\vec{\nu}}(\vec{\beta})\sum_{\substack{\lambda_i\vdash n+m\\
    \vec{\rho}\in S^d_{n+m}}}\frac{1}{(n+m)!^d}\chi_{\vec{\lambda}}(\vec{\alpha}\circ \vec{\beta})\chi_{\vec{\lambda}}(\vec{\rho})\mathcal{O}_{\vec{\rho}}\nonumber \\
    &=&\frac{d_{\vec{\mu}}d_{\vec{\nu}}}{(n!m!)^{d-1}}\sum_{\vec{\alpha}\in S_n^d}\sum_{\vec{\beta}\in S_m^d}\chi_{\vec{\mu}}(\vec{\alpha})\chi_{\vec{\nu}}(\vec{\beta})\sum_{\lambda_i\vdash n+m}\frac{1}{(n+m)!d_{\vec{\lambda}}}\chi_{\vec{\lambda}}(\vec{\alpha}\circ \vec{\beta})\mathcal{O}_{\vec{\lambda}}\nonumber \\
 &=&\frac{n!m!}{(n+m)!}\sum_{\lambda_i\vdash n+m}\frac{d_{\vec{\mu}}d_{\vec{\nu}}}{d_{\vec{\lambda}}}\,C^{\vec{\lambda}}_{\vec{\mu} \vec{\nu}}\,\mathcal{O}_{\vec{\lambda}},
\end{eqnarray}
where
\begin{equation}
C^{\vec{\lambda}}_{\vec{\mu} \vec{\nu}}=C_{\mu_1\nu_1}^{\lambda_1}\cdots C_{\mu_d\nu_d}^{\lambda_d},\qquad \mu_i\vdash m,\quad \nu_i\vdash n,\quad \lambda_i\vdash n+m.
\end{equation}

\section{Casimir operators acting on irreducible representations}
In order to compute the action of cut operators on Schur invariants $\mathcal{O}_{\vec{\mu}},$ it is necessary to know how certain Casimir operators of the group algebra $\mathbb{C}(S_{n})$ act on representations.
It is known that when a Casimir operator\footnote{Casimir operators are elements which belong to the center of $\mathbb{C}(S_{n})$, that is, they commute with every element of $\mathbb{C}(S_{n}).$} acts on an irreducible representation of $S_n$ results in a multiple of the identity of that irreducible representation. That is
\begin{equation}\label{casimirgen}
    \Gamma_{\mu}(\mathcal{C}\,\sigma )=C(\mu)\Gamma_{\mu}(\sigma),\quad \mu\vdash n,\quad \sigma\in S_n,
\end{equation}
where $\mathcal{C}$ is a Casimir of $\mathbb{C}(S_{n})$ and $C(\mu)$ is a number. Let us see some examples. Consider the Casimir $\mathbb{C}(S_{n})$ built as a sum of all transpositions,
\begin{equation}
    \mathcal{T}_2(n)=\sum_{i<j}^n(i\,j).
\end{equation}
Now, according to \eqref{casimirgen},
\begin{equation}\label{casimirtrans}
     \Gamma_{\mu}(\mathcal{T}_2(n)\,\sigma )=T_2(\mu)\Gamma_{\mu}(\sigma),\quad \mu\vdash n,\quad \sigma\in S_n. 
\end{equation}
In order to find out the value of $T_2(\mu)$, we make $\sigma=1$ and take traces in \eqref{casimirtrans}. So,
\begin{equation}\label{charactert2}
    \chi_{\mu}(\mathcal{T}_2(n))=T_2(\mu)\chi_{\mu}(1)=T_2(\mu)d_\mu.
\end{equation}
The LHS of \eqref{charactert2} is the sum of ${n}\choose{2}$ characters of $\Gamma_\mu$ evaluated on a transposition. Using the Murnaghan–Nakayama rule it is easy to find that the character of any two-cycle element
is 
\begin{equation}\label{charactertranspo}
    \chi_{\mu}((i\,j))=\frac{d_{\mu}}{\binom{n}{2} }\sum_{i=1}^{l(\mu)}\Big(\frac{\mu_i(\mu_i-1)}{2}-\frac{\mu^t_i(\mu^t_i-1)}{2}\Big),
\end{equation}
where $\mu_i$ is the length of row $i$ and the superscript $t$ stands indicates the transposed diagram. Inserting  \eqref{charactertranspo} into \eqref{charactert2} we find
\begin{equation}\label{T2mu}
  T_2(\mu)=  \sum_{i=1}^{l(\mu)}\Big(\frac{\mu_i(\mu_i-1)}{2}-\frac{\mu^t_i(\mu^t_i-1)}{2}\Big).
\end{equation}
For our purposes it will be useful to find the action of the Jucys-Murphy element 
\begin{equation}
    \mathcal{J}_n=(n\,1)+(n\,2)+\dots (n\,n-1),
\end{equation}
which is a Casimir of $\mathbb{C}(S_{n-1})$, onto irreducible representations of $S_{n-1}$. Specifically, we will need to compute $\chi_\mu(\sigma'\mathcal{J}_n)$, where $\mu\vdash n$, but $\sigma'\in S_{n-1}$. First we realize that the Jucys-Murphy element is a sum of two Casimir operators
\begin{equation}\label{JninTn}
 \mathcal{J}_n=  \mathcal{T}_2(n)-\mathcal{T}_2(n-1).
\end{equation}
Both $\mathcal{T}_2(n)$ and $\mathcal{T}_2(n-1)$ commute with all the elements of $\mathbb{C}(S_{n-1})$. We compute
\begin{equation}
     \chi_{\mu}(\mathcal{J}_n\,\sigma' )=\chi_{\mu}(\mathcal{T}_2(n)\,\sigma' )-\chi_{\mu}(\mathcal{T}_2(n-1)\,\sigma' ).
\end{equation}
Remember that if $\sigma'\in S_{n-1}$,
\begin{equation}
    \Gamma_{\mu}(\sigma')=\bigoplus_{\mu'\nearrow \mu}\Gamma_{\mu'}(\sigma').
\end{equation}
The operator $\mathcal{T}_2(n-1)$ acts on each irreducible representation of the direct sum. Applying  \eqref{T2mu} we obtain
\begin{equation}\label{contentbox}
   \boxed{ \chi_{\mu}(\mathcal{J}_n\,\sigma' )=\sum_{\mu'\nearrow \mu}J(\mu,\mu')\chi_{\mu'}(\sigma'),} 
\end{equation}
where $J(\mu,\mu')$ is the content of the (corner) box which must be deleted from diagram $\mu$ to obtain diagram $\mu'$. The content of the box in position $(i,j)$ (that is, the box at row $i$ and column $j$) is simply $j-i$. An example of a Young diagram where the content of the boxes have been spelled out is
\begin{equation}
 \begin{Young}
$0$&1&2&3&4\cr
-1&0&1&2\cr
-2&-1&0\cr
-3\cr
\end{Young}.   
\end{equation}

\section{Casimir operators on the diagonal action}
With the notation we have been using, a diagonal action on a product of irreducible representations is defined as
\begin{equation}
    \sigma\cdot \Gamma_{\vec{\mu}}(\vec{\alpha})=\Gamma_{\vec{\mu}}(\vec{\alpha}\cdot \sigma).
\end{equation}
As usual, when an action is defined on a vector space it automatically splits the space into subspaces, which are irreducible representations of the group. The quantities that appear in this paper involve sums like
\begin{equation}\label{diagonalsumrep}
  \sum_{\sigma\in S_n}\Gamma_{\vec{\mu}}(\vec{\alpha}\cdot \sigma).  
\end{equation}
The object \eqref{diagonalsumrep} is indeed an irreducible representation of the diagonal action, specifically it is the symmetric representation $(n)$ of the diagonal action. As in the case of non-diagonal actions, a Casimir operator acting on it will result in a multiple of the identity. For instance,
\begin{equation}
      \sum_{\sigma\in S_n}\Gamma_{\vec{\mu}}(\vec{\alpha}\cdot \sigma \, \mathcal{T}_2(n))= T(\vec{\mu})\sum_{\sigma\in S_n}\Gamma_{\vec{\mu}}(\vec{\alpha}\cdot \sigma).
\end{equation}
Taking traces and $\vec{\alpha}=\vec{1}$, we see that
\begin{equation}
     \sum_{\sigma\in S_n}\chi_{\vec{\mu}}(\sigma \, \mathcal{T}_2(n))=T(\vec{\mu})\sum_{\sigma\in S_n}\chi_{\vec{\mu}}(\sigma),
\end{equation}
from which we find that
\begin{equation}
    T(\vec{\mu})=\binom{n}{2}.
\end{equation}
In this paper we need to compute the more involved quantity $\sum_{\sigma'\in S_{n-1}}\Gamma_{\vec{\mu}}( \sigma' \, \mathcal{J}_n)$. As before, we will be using  the fact that
\begin{equation}
    \Gamma_{\vec{\mu}}( \sigma')=\bigoplus_{\vec{\mu}'\nearrow \vec{\mu}}\Gamma_{\vec{\mu}'}(\sigma').
\end{equation}
First, note that by just splitting the terms in the sum, we have
\begin{equation}\label{jucysdiagonal1}
   \sum_{\sigma'\in S_{n-1}}\Gamma_{\vec{\mu}}( \mathcal{J}_n\,\sigma')= \sum_{\sigma\in S_{n}}\Gamma_{\vec{\mu}}( \sigma)-\sum_{\sigma'\in S_{n-1}}\Gamma_{\vec{\mu}}( \sigma').
\end{equation}
Now, since $\mathcal{J}_n$ is a Casimir of $\mathbb{C}(S_{n-1})$, its action on an irreducible representation is proportional to the identity, so
\begin{equation}\label{jucysdiagonal2}
   \boxed{\sum_{\sigma'\in S_{n-1}}\Gamma_{\vec{\mu}}( \mathcal{J}_n\,\sigma')= J(\vec{\mu})\sum_{\sigma'\in S_{n-1}}\Gamma_{\vec{\mu}}( \sigma').} 
\end{equation}
Taking traces in \eqref{jucysdiagonal1} and in \eqref{jucysdiagonal2}, we find 
\begin{equation}
    \big(J(\vec{\mu})+1\big)\sum_{\sigma'\in S_{n-1}}\chi_{\vec{\mu}}( \sigma')= \sum_{\sigma\in S_{n}}\chi_{\vec{\mu}}( \sigma).
\end{equation}
Let us call
\begin{equation}
    g_{\vec{\mu}'\uparrow \vec{\mu}}\equiv \sum_{\vec{\mu}'\nearrow\vec{\mu}} g_{\vec{\mu}'}=\frac{1}{(n-1)!}\sum_{\sigma'\in S_{n-1}}\chi_{\vec{\mu}}(\sigma').
\end{equation}
Then
\begin{equation}
  \sum_{\sigma'\in S_{n-1}}\chi_{\vec{\mu}}( \sigma')= \sum_{\substack{\sigma'\in S_{n-1} \\
  \vec{\mu}'\nearrow\vec{\mu}}}\chi_{\vec{\mu}'}( \sigma')=(n-1)! \,g_{\vec{\mu}'\uparrow \vec{\mu}},
\end{equation}
and 
\begin{equation}\label{jucysconstant}
 \boxed{ J(\vec{\mu})=n\frac{g_{\vec{\mu}}}{g_{\vec{\mu}'\uparrow \vec{\mu}}}-1.} 
\end{equation}
It is easy to prove that \eqref{jucysconstant} agrees with the prescription given above for the case of only one representation,
\begin{equation}
 \sum_{\sigma'\in S_{n-1}}\Gamma_{\mu}( \mathcal{J}_n\,\sigma')=J(\mu) \sum_{\sigma'\in S_{n-1}}\Gamma_{\mu}( \sigma'),   
\end{equation}
where $g_\mu=\delta_{\mu (n)}$ and  $g_{\mu'\uparrow \mu}=1$. In this case
\begin{equation}
   J\big(\mu=(n)\big)=n-1,\qquad J\big(\mu=(n-1,1)\big)=-1,
\end{equation}
being zero in the rest of the cases,
as the prescription of the content of the boxes indicates. \\
Using \eqref{jucysconstant}, we see from \eqref{jucysdiagonal1} that
\begin{equation}\label{result}
   \boxed{ \sum_{\sigma\in S_{n}}\Gamma_{\vec{\mu}}( \sigma)=n\frac{g_{\vec{\mu}}}{g_{\vec{\mu}'\uparrow \vec{\mu}}}\sum_{\sigma'\in S_{n-1}}\Gamma_{\vec{\mu}}( \sigma'),}
\end{equation}
which is the result we are using in appendix \ref{CAODS}.

\section{Cut operators acting on Schur invariants}\label{CAODS}
The cut operator defined in \cite{ito2} reads
\begin{equation}
\Delta \mathcal{O}= \frac{\delta^2 \mathcal{O}[T]}{\delta T_{i_1\dots i_d} \delta \overline{T}^{i_1\dots i_d}}.
\end{equation}
In this appendix, to simplify notation, we are going to consider tensors with three indices, what means that $d=3$ in what follows. The general case can be straightforwardly recovered.\\
The cut operator acts on Schur invariants as
\begin{eqnarray}\label{doublevariation}
 \Delta \mathcal{O}_{\vec{\mu}}&=&\frac{1}{n!^d}\sum_{\substack{r,s=1\\
 \alpha,\alpha_2,\alpha_3,\sigma\in S_n}}^{n}d_{\vec{\mu}}\chi_{\vec{\mu}}(\vec{\alpha}\cdot\sigma)\delta_{i_r}^{i_{\alpha_1(s)}}\delta_{j_r}^{j_{\alpha_2(s)}}\delta_{k_r}^{k_{\alpha_3(s)}} \times \nonumber \\
&&T_{i_1j_1k_1}\underset{\underset{r}{\wedge}}{\cdots} T_{i_nj_nk_n}\overline{T}^{i_{\alpha_1(1)}j_{\alpha_2(1)}k_{\alpha_3(1)}}\underset{\underset{s}{\wedge}}{\cdots} \overline{T}^{i_{\alpha(n)}j_{\alpha_2(n)}k_{\alpha_3(n)}}.
\end{eqnarray}
Performing the changes
\begin{equation}
\tilde{\alpha}_i=\big(n~r\big)\, \alpha_i\, \big(n~s\big), 
\end{equation}
taking into account that 
\begin{equation}
\sum_{\sigma\in S_n} \chi_{\vec{\mu}}(\vec{\tilde{\alpha}}\cdot\sigma)=\sum_{\sigma\in S_n} \chi_{\vec{\mu}}(\vec{\alpha}\cdot\sigma),   
\end{equation}
and relabeling the permutations, we arrive at
\begin{eqnarray}\label{doublevariationcont}
 \Delta \mathcal{O}_{\vec{\mu}}&=&\frac{n^2}{n!^d}\sum_{
 \vec{\alpha},\sigma}d_{\vec{\mu}}\chi_{\vec{\mu}}(\vec{\alpha}\cdot\sigma)\delta_{i_n}^{i_{\alpha_1(n)}}\delta_{j_n}^{j_{\alpha_2(n)}}\delta_{k_n}^{k_{\alpha_3(n)}}\times \nonumber \\
&&T_{i_1j_1k_1}\cdots T_{i_{n-1}j_{n-1}k_{n-1}}\overline{T}^{i_{\alpha_1(1)}j_{\alpha_2(1)}k_{\alpha_3(1)}}\cdots \overline{T}^{i_{\alpha_1(n-1)}j_{\alpha_2(n-1)}k_{\alpha_3(n-1)}}.
\end{eqnarray}
First, we apply the result \eqref{result} in the appendix. The diagonal sum over $\sigma\in S_n$ turns into a sum over $\sigma'\in S_{n-1}$ producing  a global factor. We obtain
\begin{eqnarray}\label{doublevariationcont1}
 \Delta \mathcal{O}_{\vec{\mu}}&=&\frac{n^3}{n!^d}\frac{g_{\vec{\mu}}}{g_{\vec{\mu}'\uparrow \vec{\mu}}}\sum_{
 \vec{\alpha},\sigma'}d_{\vec{\mu}}\chi_{\vec{\mu}}(\vec{\alpha}\cdot\sigma')\delta_{i_n}^{i_{\alpha_1(n)}}\delta_{j_n}^{j_{\alpha_2(n)}}\delta_{k_n}^{k_{\alpha_3(n)}}\times \nonumber \\
&&T_{i_1j_1k_1}\cdots T_{i_{n-1}j_{n-1}k_{n-1}}\overline{T}^{i_{\alpha_1(1)}j_{\alpha_2(1)}k_{\alpha_3(1)}}\cdots \overline{T}^{i_{\alpha_1(n-1)}j_{\alpha_2(n-1)}k_{\alpha_3(n-1)}}.
\end{eqnarray}
The next step is to write the elements  $\alpha_1,\alpha_2,\alpha_3 \in S_n$ as elements of $S_{n-1}$ composed with a transposition. Note that the decomposition
\begin{equation}\label{parametrization}
\alpha=(s\,n)\alpha',\quad \alpha\in S_n,\quad \alpha'\in S_{n-1}, \quad s=1,\dots,n,
\end{equation}
where $\alpha'$ does not involve $n$, is unique. All permutations $\alpha\in S_n$ are obtained without repetition as we run over $\alpha'\in S_{n-1}$ and $s=1,\dots, n$ in \eqref{parametrization}.  With the parameterization \eqref{parametrization} $\alpha(n)=s$, and $\alpha(n)=n$ only when $s=n$. Besides, if written in disjoint cycles notation,  the permutation $\alpha'$ is obtained from $\alpha$ by simply deleting the ``letter'' $n$. \\
Now, Let us decompose each sum over $\alpha_i$ in \eqref{doublevariationcont1} into the sum over $\alpha'_i$ and the sum over $\mathcal{J}_n\alpha'_i$. This splitting has a purpose. Note that 
\begin{equation*}
    \delta_{i_n}^{i_{\alpha_1(n)}}\delta_{j_n}^{j_{\alpha_2(n)}}\delta_{k_n}^{k_{\alpha_3(n)}} T_{i_1j_1k_1}\cdots T_{i_{n-1}j_{n-1}k_{n-1}}\overline{T}^{i_{\alpha_1(1)}j_{\alpha_2(1)}k_{\alpha_3(1)}}\cdots \overline{T}^{i_{\alpha_1(n-1)}j_{\alpha_2(n-1)}k_{\alpha_3(n-1)}}=N^a\mathcal{O}_{\vec{\alpha}'},
\end{equation*}
where $a$ is the number of $\alpha$'s for which $\alpha_i(n)=n$. This is implemented in \eqref{doublevariationcont1} as
\begin{equation}
    \Delta \mathcal{O}_{\vec{\mu}}=\frac{n^3}{n!^d}\frac{g_{\vec{\mu}}}{g_{\vec{\mu}'\uparrow \vec{\mu}}}\sum_{
 \vec{\alpha}',\sigma'}d_{\vec{\mu}}\prod_{k=1}^3\big[N\chi_{\mu_k}(\alpha'_k\sigma')+\chi_{\mu_k}(\mathcal{J}_n\alpha'_k\sigma')\big]\mathcal{O}_{\alpha'}.
\end{equation}
Applying \eqref{contentbox} and the fact that 
\begin{equation}
    \chi_{\mu}(\alpha')=\sum_{\mu'\nearrow \mu}\chi_{\mu'}(\alpha'),
\end{equation}
we can write
\begin{eqnarray}\label{finalresult}
    \Delta \mathcal{O}_{\vec{\mu}}&=&\frac{n^3}{n!^d}\frac{g_{\vec{\mu}}}{g_{\vec{\mu}'\uparrow \vec{\mu}}}\sum_{
 \vec{\alpha}',\sigma'}\sum_{\vec{\mu}'\nearrow \vec{\mu}}d_{\vec{\mu}}\chi_{\vec{\mu}'}(\vec{\alpha}'\sigma')\prod_{k=1}^3\big[N+J(\mu_k,\mu'_k)\big]\mathcal{O}_{\vec{\alpha}'} \nonumber \\
 &=&\frac{n^3 (n-1)!^d}{n!^d}\frac{g_{\vec{\mu}}}{g_{\vec{\mu}'\uparrow \vec{\mu}}}\sum_{\vec{\mu}'\nearrow \vec{\mu}}\frac{d_{\vec{\mu}}}{d_{\vec{\mu}'}}\prod_{k=1}^3\big[N+J(\mu_k,\mu'_k)\big]\mathcal{O}_{\vec{\mu}'} \nonumber \\
 &=&n^3\frac{g_{\vec{\mu}}}{g_{\vec{\mu}'\uparrow \vec{\mu}}}\sum_{\vec{\mu}'\nearrow \vec{\mu}}\frac{\text{Dim}_{\vec{\mu}}(N)}{\text{Dim}_{\vec{\mu}'}(N)}\mathcal{O}_{\vec{\mu}'},
\end{eqnarray}
where, in the last line of \eqref{finalresult}, we have applied
\begin{equation}
    N+J(\mu_k,\mu'_k)=\frac{f_{\mu_k}}{f_{\mu'_k}}=n\frac{\text{Dim}_{\vec{\mu}}(N)}{\text{Dim}_{\vec{\mu}'}(N)}\frac{d_{\mu'_k}}{d_{\mu_k}}.
\end{equation}
Note that the factor $n^3$ in the last line of \eqref{finalresult} is general, valid for tensors of any order. The result \eqref{finalresult} is a proof that the cut operator is closed when acting on Schur operators $\mathcal{O}_{\vec{\mu}}$.

\end{document}